\documentclass[english,aps, reprint, prb, amsmath,amssymb,floatfix,superscriptaddress,eqsecnum,longbibliography]{revtex4-2}
\usepackage[T1]{fontenc}
\usepackage[latin9]{inputenc}
\usepackage{babel}
\usepackage{amsmath}
\usepackage{amssymb}
\usepackage{graphicx}
\usepackage[unicode=true,pdfusetitle,
 bookmarks=true,bookmarksnumbered=false,bookmarksopen=false,
 breaklinks=false,pdfborder={0 0 1},backref=false,colorlinks=false]
 {hyperref}
\usepackage{xcolor}

\def\o{\ell}

\makeatletter
\usepackage{braket}

\makeatother

\begin{document}
\title{Exchange constants for local spin Hamiltonians from tight-binding models}
\author{Simon Streib}
\affiliation{Department of Physics and Astronomy, Uppsala University, Box 516,
SE-75120 Uppsala, Sweden}
\author{Attila Szilva}
\affiliation{Department of Physics and Astronomy, Uppsala University, Box 516,
SE-75120 Uppsala, Sweden}
\author{Vladislav Borisov}
\affiliation{Department of Physics and Astronomy, Uppsala University, Box 516,
SE-75120 Uppsala, Sweden}
\author{Manuel Pereiro }
\affiliation{Department of Physics and Astronomy, Uppsala University, Box 516,
SE-75120 Uppsala, Sweden}
\author{Anders Bergman}
\affiliation{Department of Physics and Astronomy, Uppsala University, Box 516,
SE-75120 Uppsala, Sweden}
\author{Erik Sj\"oqvist }
\affiliation{Department of Physics and Astronomy, Uppsala University, Box 516,
SE-75120 Uppsala, Sweden}
\author{Anna Delin }
\affiliation{Department of Applied Physics, School of Engineering Sciences, KTH
Royal Institute of Technology, Electrum 229, SE-16440 Kista, Sweden}
\affiliation{Swedish e-Science Research Center (SeRC), KTH Royal Institute of Technology, SE-10044 Stockholm, Sweden}
\author{Mikhail I. Katsnelson}
\affiliation{Institute for Molecules and Materials, Radboud University, Heyendaalseweg 135, 6525 AJ, Nijmegen, The Netherlands}
\author{Olle Eriksson }
\affiliation{Department of Physics and Astronomy, Uppsala University, Box 516,
SE-75120 Uppsala, Sweden}
\affiliation{School of Science and Technology, \"Orebro University, Sweden}
\author{Danny Thonig}
\affiliation{School of Science and Technology, \"Orebro University, Sweden}
\affiliation{Department of Physics and Astronomy, Uppsala University, Box 516,
SE-75120 Uppsala, Sweden}
\date{May 26, 2021}
\begin{abstract}
We consider the mapping of tight-binding electronic structure theory to a local spin Hamiltonian,
based on the adiabatic approximation for spin degrees of freedom in itinerant-electron systems. Local spin Hamiltonians are introduced in order to describe the energy landscape of
small magnetic fluctuations, locally around a given spin configuration. They are designed for linear response near a given magnetic state and in general insufficient to capture
arbitrarily strong deviations of spin configurations from the equilibrium. In order to achieve this mapping, we include a linear term in the local spin Hamiltonian that, together with the usual bilinear exchange tensor, produces an improved accuracy of effective magnetic Weiss fields for non-collinear states. We also provide examples from tight-binding electronic structure theory, where our implementation of the calculation of exchange constants is based on constraining fields
that stabilize an out-of-equilibrium spin configuration. We check our formalism by means of numerical calculations for iron dimers and chains.
\end{abstract}
\maketitle

\section{Introduction}

The Heisenberg model and generalizations thereof are among the most
important paradigms of condensed matter physics and have been very
successful in describing the magnetic behavior of both magnetic insulators for which it was suggested initially and, with some reservations, also metallic magnets. There are several complementary approaches
of obtaining the exchange parameters that enter an effective atomistic spin
Hamiltonian for a specific material: (a) one can obtain the exchange
parameters analytically from a more fundamental electronic Hamiltonian
\citep{Takahashi1977,MacDonald1988,Hoffmann2020}, (b) one can map the spin Hamiltonian onto total
energy calculations for spin spirals \citep{Kuebler} and spin-cluster expansions \citep{Drautz2004,Drautz2005,Antal2008, Grytsiuk2020,Brinker2020},
or (c) one can use energy variations of the magnetic ground state within first-principle approaches such as spin-density functional theory
\citep{Liechtenstein1984,Liechtenstein1987}. While approaches (a)
and (b) aim to describe arbitrary spin configurations, approach (c)
is explicitly designed to capture small fluctuations around the magnetic
ground state. In this paper, we will focus on approach (c), which
was pioneered by Liechtenstein, Katsnelson, Antropov, and Gubanov
(LKAG) \citep{Liechtenstein1984,Liechtenstein1987}. 

In their work, Liechtenstein \emph{et al. }emphasize that
for metals the Heisenberg Hamiltonian \emph{``is applicable only
for small spin deviations from the ground state}'' \citep{Liechtenstein1987},
which implies that terms beyond the bilinear Heisenberg exchange interactions
may be required to describe the magnetic behavior for strong deviations
from the ground state. This was further confirmed by explicit calculations for the cases of Fe, Ni, and Fe-based magnetic alloys \citep{Turzhevskii1990}. While the original work by Liechtenstein \emph{et
al. }considered ferromagnetic ground states, extensions of the LKAG
formalism to non-equilibrium \citep{Secchi2013,Secchi2016} and non-collinear \citep{Antropov1997,Antropov1999,Szilva2013,Secchi2015,Szilva2017,Cardias2020a,Cardias2020b} states have been considered.
However, it was realized that a mapping of non-collinear spin configurations
to a Heisenberg model \citep{Szilva2017} or to a generalized Heisenberg
model with a bilinear exchange tensor \citep{Szilva2013} is in general
not possible for non-collinear states, apparently requiring the inclusion
of higher order (beyond Heisenberg) exchange contributions \citep{Hoffmann2020,Brinker2020,Dias2021}.
We propose an alternative solution by including a linear term in a
generalized spin model. Linear terms are usually not considered in effective spin Hamiltonians due to arguments connected to degeneracy of time-reversed states. We argue here that such a linear term can be considered if one is interested only in small
fluctuations around a given spin configuration and takes into account that the sign of the linear interaction parameter changes for the spin-reversed configuration. As outlined here, higher order exchange
interactions may not be required in this case, which is in line with
the original LKAG approach \citep{Liechtenstein1984,Liechtenstein1987}.

It is well established that the LKAG formalism is only exact in the
long-wavelength regime \citep{Bruno2003,Antropov2003,Katsnelson2004}.
In an implementation of the adiabatic approximation beyond the
long-wavelength limit, the inclusion of constraining fields is required
\citep{Stocks1998,Ujfalussy1999,Bruno2003}. These constraining fields
stabilize non-collinear, out-of-equilibrium spin configurations. We
present a formalism for calculating the full bilinear exchange tensor from
tight-binding models, which is based on the fact that the effective
magnetic field is the negative of the constraining field \citep{Stocks1998,Ujfalussy1999,Streib2020,Cardias2021}. {This follows from the physical picture that the constraining field has to cancel out the effective field acting on a spin.}
From recent results that the effective field in density functional theory (DFT) contains an additional
term besides the constraining field \citep{Streib2020}, we recover
the formula previously derived by Bruno {for the isotropic Heisenberg exchange \citep{Bruno2003}, which we extend to the full bilinear exchange tensor}. Results
for the exchange parameters of nickel from a formalism similar to
Ref.~\citep{Bruno2003} have been recently published \citep{Solovyev2020}
and show a similar behavior as previous results based on a frozen
magnon approach \citep{Jacobsson2017}, which also takes constraining
fields into account. Note that, as was shown analytically \citep{Katsnelson2004},
the Bruno formula corresponds to extraction of exchange parameters from the
inverse static magnetic susceptibility, that is, to the energy of static spiral configurations, whereas LKAG exchanges correspond to the
poles of dynamic magnetic susceptibility, that is, to the spin-wave spectrum measured, e.g., by inelastic neutron scattering. In the formal limit of well-defined local magnetic moments where intersite exchange energies are much smaller than on-site Hund exchange splitting these two expressions coincide. 

The paper is organized as follows: in Sec.~\ref{sec:hamiltonians}
we introduce and define the local spin Hamiltonian and derive an explicit
expression for the linear and bilinear terms from the effective magnetic
field. In Sec.~\ref{sec:exchange} we derive expressions based on the constraining field for the exchange parameters in terms of Green's functions and self-energies. We apply these formulas
in Sec.~\ref{sec:numerics} to a tight-binding model for iron and
present results for iron dimers and chains. Finally in Sec.~\ref{sec:summary}
we summarize our results and provide a broader contextual analysis. In the Appendices
\ref{sec:projection} and \ref{sec:matsubara}, we give additional
details on the definition of the effective field and the calculation
of Matsubara sums, respectively. {In Appendix \ref{sec:symmetry}, we discuss the symmetry of exchange constants within the different approaches that we consider in this manuscript.}

\section{Local spin Hamiltonians\label{sec:hamiltonians}}

\begin{figure}
\begin{centering}
\includegraphics[width=0.9\columnwidth]{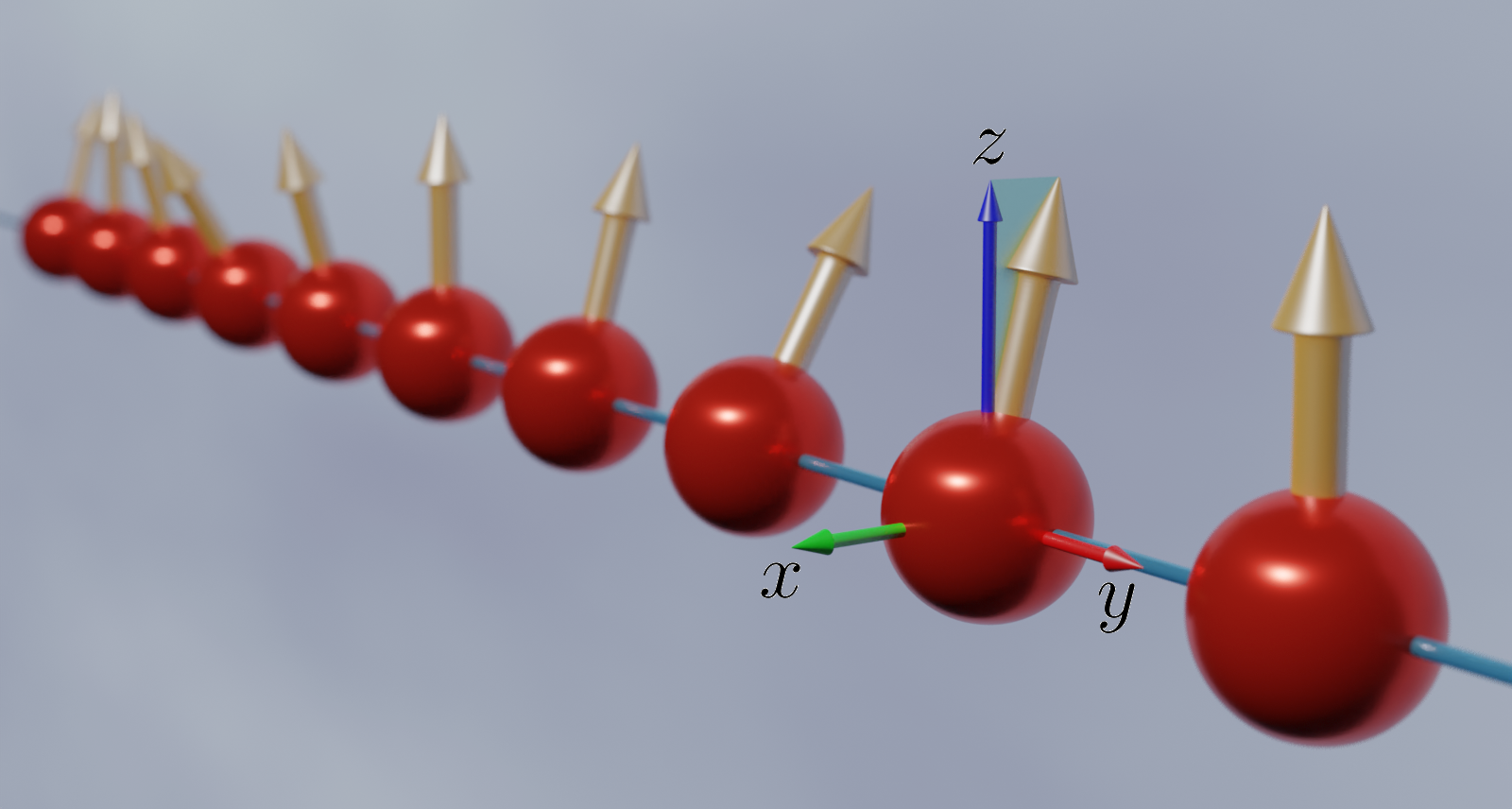}
\par\end{centering}
\caption{Illustration of a non-collinear spin chain studied as an example in Sec.~\ref{sec:hamiltonians}, 
including the definition of the reference frame.\label{fig:spin_chain}} 
\end{figure}

We distinguish between two types of spin Hamiltonians: global and
local Hamiltonians. With global we denote a Hamiltonian that aims
to describe energies of all possible spin configurations (approaches (a) and (b)
above), while a local Hamiltonian is designed to describe energetics of spin configurations
in the vicinity of the ground state or, more generally, in the vicinity
of a predefined spin configuration (approach (c) above). Global Hamiltonians
are in principle superior, but in practice it may be difficult to
obtain the necessary parameters for higher order exchange interactions if they play a significant role. Also, rigorously speaking, it is not guaranteed that the global Hamiltonian, expressed in terms of spin operators only, exists for itinerant-electron systems at all. Local spin Hamiltonians do not require
any spin interactions beyond the bilinear order (for Heisenberg exchange as well as Dzyaloshinskii-Moriya interactions) and the exchange parameters
can be directly computed without the need to fit the spin Hamiltonian
to many different spin configurations. {However, for a given local spin Hamiltonian, the range of validity, i.e. how small the fluctuations should be, is a priori not known and depends on how significant higher-order exchange contributions are.} Thus, local and global spin
Hamiltonians are complementary approaches with distinct advantages
and disadvantages. To avoid further misunderstanding we have to emphasize once more that we mean here ``locality'' and ``globality'' of the Hamiltonians in energy and not in real space. 

In its most general form, the local spin Hamiltonian we consider in this work is defined as
\begin{equation}
\mathcal{H}_{s}=-\sum_{i\alpha}C_{i\alpha}e_{i\alpha}-\frac{1}{2}\sum_{ij}\sum_{\alpha\beta}J_{ij}^{\alpha\beta}e_{i\alpha}e_{j\beta},\label{eq:local Hamiltonian}
\end{equation}
where $e_{i\alpha}$ is the component $\alpha=x,y,z$ of the magnetic
moment unit vector at site $i$, $J_{ij}^{\alpha\beta}$ is the exchange
tensor, and we allow for a linear contribution, $C_{i\alpha}$. Note that in this formulation the size of the magnetic moment on each site is incorporated in the value of $J_{ij}^{\alpha\beta}$.
The linear term in Eq.~(\ref{eq:local Hamiltonian}) is an important difference between the local and global
approaches. In the global approach a linear term is not allowed because
the Hamiltonian (without an external magnetic field) has to be invariant under an inversion of all magnetic
moment directions, $e_{i\alpha}\to-e_{i\alpha}$, due to time-reversal
symmetry. For the local approach the linear term is allowed since
only small fluctuations are described and the inversion of all moment
directions is beyond this regime. Time-reversal symmetry is recovered
by considering that the parameter $C_{i\alpha}$ changes its sign
for a time-reversed reference state.

The effective Weiss field of the Hamiltonian (\ref{eq:local Hamiltonian}), which is relevant for spin dynamics and for obtaining an equilibrium configuration of the atomic moments, 
is given by
\begin{align}
B_{i\alpha}^{\text{eff}} & =-\frac{1}{M_{i}}\frac{\partial\mathcal{H}_{s}}{\partial e_{i\alpha}}=\frac{1}{M_{i}}C_{i\alpha}+\frac{1}{M_{i}}\sum_{k\nu}J_{ik}^{\alpha\nu}e_{k\nu},\label{eq:effective field}
\end{align}
where $M_{i}$ is the magnetic moment length at site $i$. To specify
the parameters of the spin Hamiltonian, we consider the following
expansion of the effective field around a given spin configuration
$\{\mathbf{e}_{i}^{0}\}$ to the first order,
\begin{equation}
B_{i\alpha}^{\text{eff}}\approx B_{i\alpha}^{\text{eff}}(\{\mathbf{e}_{i}^{0}\})+\sum_{j\beta}\left.\frac{\partial B_{i\alpha}^{\text{eff}}}{\partial e_{j\beta}}\right|_{\{\mathbf{e}_{i}^{0}\}}\left(e_{j\beta}-e_{j\beta}^{0}\right).\label{eq:effective field expansion}
\end{equation}
By comparing Eqs.~(\ref{eq:effective field}) and (\ref{eq:effective field expansion}),
we obtain
\begin{align}
J_{ij}^{\alpha\beta} & =M_{i}\left.\frac{\partial B_{i\alpha}^{\text{eff}}}{\partial e_{j\beta}}\right|_{\{\mathbf{e}_{i}^{0}\}},\\
C_{i\alpha} & =M_{i}B_{i\alpha}^{\text{eff}}(\{\mathbf{e}_{i}^{0}\})-\sum_{k\nu}J_{ik}^{\alpha\nu}e_{k\nu}^{0}.\label{eq:linear term}
\end{align}
As required by time-reversal symmetry,
\begin{equation}
C_{i\alpha}(\{\mathbf{e}_{i}^{0}\})=-C_{i\alpha}(\{-\mathbf{e}_{i}^{0}\}).
\end{equation}

If the system under consideration can be exactly described by a bilinear
spin Hamiltonian without any higher order terms, we have
\begin{equation}
B_{i\alpha}^{\text{eff}}(\{\mathbf{e}_{i}^{0}\})=\frac{1}{M_{i}}\sum_{k\nu}J_{ik}^{\alpha\nu}e_{k\nu}^{0},
\end{equation}
and the parameter $C_{i\alpha}$ vanishes and is not required. In
that case the local and global spin Hamiltonians are identical. The
linear term in the local Hamiltonian plays therefore only a role if
higher order exchange interactions are present in the global Hamiltonian.

{
In the ground state, $\mathbf{B}_{i}^\mathrm{eff}=-\boldsymbol{\nabla}_{\mathbf{e}_i}E/M_i$ vanishes in Eq.~(\ref{eq:linear term}), since the gradient of the energy $E$ has to be zero, and $C_{i\alpha}$ is then determined by the exchange tensor $J_{ij}^{\alpha\beta}$ alone. We note that for the effective magnetic field that drives the precession term of the dynamics of
the moment direction $\mathbf{e}_{i}$, 
\begin{equation}
\dot{\mathbf{e}}_{i}=\gamma\mathbf{e}_{i}\times\mathbf{B}_{i}^{\text{eff}},\label{eq:EOM}
\end{equation}
with $\gamma$ the gyromagnetic ratio, only the component of the effective field that is perpendicular to $\mathbf{e}_{i}$, i.e. $\mathbf{B}_{i\perp}^{\text{eff}}$, contributes due to the cross product.
}

For a Heisenberg model with ferromagnetic ground state aligned along the $z$ axis and isotropic exchange tensor,
\begin{equation}
    J_{ij}^{\alpha\beta}=J_{ij} \delta_{\alpha\beta},
\end{equation}
we have
\begin{align}
C_{ix} & =-\sum_{j}J_{ij}^{xz}=0,\\
C_{iy} & =-\sum_{j}J_{ij}^{yz}=0,\\
C_{iz} & =-\sum_{j}J_{ij}^{zz}=0.
\end{align}
The requirement $J_{ij}^{zz}=0$ follows from the projection to perpendicular
effective fields (see Appendix \ref{sec:projection}). Since $C_{iz}$ is the component parallel to the moment direction, it does not contribute to $\mathbf{B}_{i\perp}^{\text{eff}}$
when considering small fluctuations around the ferromagnetic ground state even if we do not consider the projection to perpendicular fields. Therefore, we recover the established result that no linear terms are required for a ferromagnetic Heisenberg model within the LKAG approach \citep{Liechtenstein1984,Liechtenstein1987}.

\begin{figure}
\begin{centering}
\includegraphics{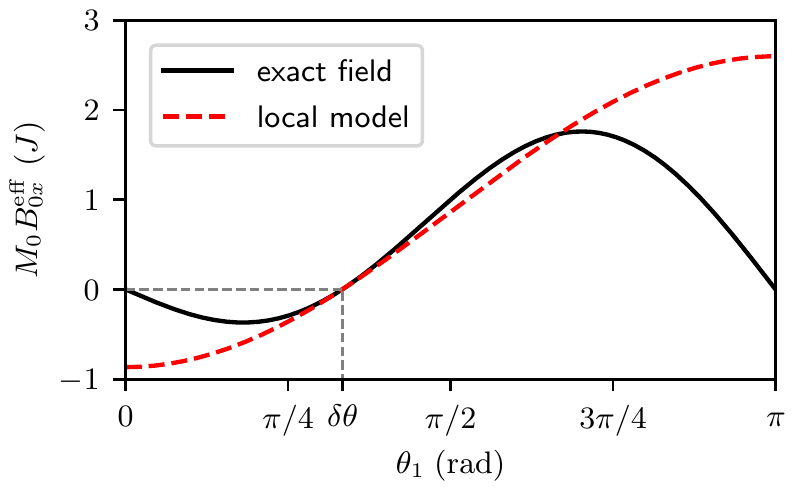}
\par\end{centering}
\caption{Comparison of the effective field for the toy model, Eq.~(\ref{eq:toy model effective field})
with $J/B=1$, and the corresponding local spin Hamiltonian, Eq.~(\ref{eq:effective field}). We consider the effective
field acting on the spin at site $i=0$ under rotations of the spin
at site $j=1$.\label{fig:example}}
\end{figure}

As a simple example, we consider a one-dimensional spin chain (see Fig.~\ref{fig:spin_chain}) with
both bilinear and biquadratic nearest-neighbour exchange contributions that are selected to result in a
non-collinear ground state,
\begin{equation}
\mathcal{H}=-J\sum_{i}\mathbf{e}_{i}\cdot\mathbf{e}_{i+1}+B\sum_{i}\left(\mathbf{e}_{i}\cdot\mathbf{e}_{i+1}\right)^{2},\label{eq:toy model}
\end{equation}
with $J,B>0$. To obtain the ground state, it is sufficient to consider
configurations with
\begin{equation}
\mathbf{e}_{i}\cdot\mathbf{e}_{i+1}=\cos(\delta\theta),\quad\forall i,
\end{equation}
where $\delta\theta$ is the angle between two neighboring spins.
The energy is minimized for
\begin{equation}
\delta\theta=\begin{cases}
0, & J/B\geq2\\
\arccos\left(\frac{J}{2B}\right), & J/B<2
\end{cases},
\end{equation}
i.e., for $J/B<2$, the ground state is non-collinear. For simplicity, we consider here only spin configurations within the $xz$ plane, such that the spin at each site $j$ (specified as an integer) is determined by a single angle $\theta_j$ with ground state value
\begin{equation}
    \theta_j=j\delta\theta.
\end{equation}
Figure \ref{fig:example}
shows a comparison of the exact effective field, obtained from the gradient of Eq.~(\ref{eq:toy model}), 
\begin{align}
M_{i}\mathbf{B}_{i}^{\text{eff}}&=-\boldsymbol{\nabla}_{\mathbf{e}_{i}}\mathcal{H}\nonumber\\&=J\left(\mathbf{e}_{i+1}+\mathbf{e}_{i-1}\right)\nonumber\\
&-2B\left(\mathbf{e}_{i+1}\left[\mathbf{e}_{i}\cdot\mathbf{e}_{i+1}\right]+\mathbf{e}_{i-1}\left[\mathbf{e}_{i}\cdot\mathbf{e}_{i-1}\right]\right),\label{eq:toy model effective field}
\end{align}
for $J/B=1$ (for which $\delta\theta = \pi/3$) and the
effective field obtained from the local spin Hamiltonian, last part of Eq.~(\ref{eq:effective field}). The results are obtained 
for the spin
at site $i=0$ under rotations of the spin $j=1$ with angle $\theta_{1}$,
while all other spins are in their non-collinear ground-state configuration. The relevant ground-state effective exchange parameters are
\begin{align}
\ensuremath{J_{0,1}^{xx}} & =\ensuremath{J_{0,-1}^{xx}}=J-2B\cos(\delta\theta),\label{eq:effective exchange1}\\
J_{0,1}^{xz} & =-J_{0,-1}^{xz}=-2B\sin(\delta\theta).\label{eq:effective exchange2}
\end{align}

The effective field vanishes in the ground state, $\theta_{1}=\delta\theta$,
as expected. The linear term $C_{0x}$ vanishes in this example due to a mirror symmetry with respect to the spins at sites $-1$ and $+1$ with $\theta_{+1}=-\theta_{-1}$.
As the figure shows, the local spin Hamiltonian
provides an excellent approximation of the effective field for small
fluctuations around the ground state. The result suggests that a local spin Hamiltonian
is sufficient for the calculation of, e.g., spin-wave spectra since they only depend on
energy variations near the ground state.

\section{Exchange constants\label{sec:exchange}}

We consider here the derivation of the exchange tensor $J_{ij}^{\alpha\beta}$
of the local spin Hamiltonian (\ref{eq:local Hamiltonian}) based on a tight-binding formalism. The standard
approach is to consider variations of the electronic energy \citep{Liechtenstein1984,Liechtenstein1987},
which may be used to obtain the exchange parameters,
\begin{equation}
J_{ij}^{\alpha\beta}=-\frac{\partial^{2}\langle\hat{\mathcal{H}}\rangle}{\partial e_{i\alpha}\partial e_{j\beta}}.\label{eq:J from E}
\end{equation}
From the point of view of a global spin Hamiltonian that may contain
higher order exchange contributions, these exchange parameters $J_{ij}^{\alpha\beta}$
are not just the bilinear exchange parameters of such a global
Hamiltonian but also take higher order exchange contributions into
account, see Eqs.~(\ref{eq:effective exchange1}) and (\ref{eq:effective exchange2}), which causes a configuration dependence of $J_{ij}^{\alpha\beta}$ \citep{Turzhevskii1990,Dias2021}.

Based on the result that the effective magnetic field, when not considering DFT calculations, can just be obtained
from the constraining field \citep{Streib2020},
\begin{equation}
B_{i\alpha}^{\text{eff}}=-\frac{1}{M_{i}}\frac{\partial\langle\hat{\mathcal{H}}\rangle}{\partial e_{i\alpha}}=-B_{i\alpha}^{\text{con}},
\end{equation}
we are taking here the alternative approach of calculating the exchange
parameters from the constraining field,

\begin{equation}
J_{ij}^{\alpha\beta}=-M_{i}\frac{\partial B_{i\alpha}^{\text{con}}}{\partial e_{j\beta}},\label{eq:J from constrain}
\end{equation}
{which is equivalent to Eq.~(\ref{eq:J from E}) if} we assume $M_i=\mathrm{const}$.
The constraining field $\mathbf{B}_i^\mathrm{con}$ is {perpendicular to the moment direction $\mathbf{e}_i$ and is }required to stabilize non-equilibrium spin configurations within the adiabatic approximation \citep{Halilov1998,Stocks1998,Ujfalussy1999, Streib2020}. It is
added here to the full electronic tight-binding Hamiltonian, $\hat{\mathcal{H}}_\mathrm{tb}$,
\begin{equation}
\hat{\mathcal{H}}=\hat{\mathcal{H}}_\mathrm{tb}+\hat{\mathcal{H}}_{\text{con}},
\end{equation}
with
\begin{equation}
\hat{\mathcal{H}}_{\text{con}}=-\sum_{i}\gamma\hat{\mathbf{S}}_{i}\cdot\mathbf{B}_{i}^{\text{con}},
\end{equation}
where $\hat{\mathbf{S}}_i$ is the spin operator at lattice site $i$.

The tight-binding Hamiltonian consists of a hopping term, $\hat{\mathcal{H}}_0$, and an interaction term, $\hat{\mathcal{H}}_{\mathrm{int}}$,
\begin{equation}
    \hat{\mathcal{H}}_{\mathrm{tb}}=\hat{\mathcal{H}}_{0}+\hat{\mathcal{H}}_{\mathrm{int}}.\label{eq:H_tb}
\end{equation}
The hopping term is in second-quantization given by
\begin{equation}\label{eqn:tightbind}
    \hat{\mathcal{H}}_{0}=\sum_{i\o,j\o',\sigma}t_{i\o,j\o'}\hat{c}_{i\o\sigma}^{\dagger}\hat{c}_{j\o'\sigma},
\end{equation}
which describes the hopping of an electron from state $j\o'\sigma$ to $i\o\sigma$ with hopping amplitude $t_{i\o,j\o'}$ and creation and annihilation operators $\hat{c}_{i\o\sigma}^{\dagger}$ and $\hat{c}_{j\o'\sigma}$. The index $i\o\sigma$ indicates the lattice site, orbit, and spin, respectively. 
The hopping amplitudes are assumed to be constant parameters that do not depend on the magnetic state of the system, which is a consequence of expressing the tight-binding Hamiltonian in Eq.~(\ref{eqn:tightbind}) in a global spin-basis (defined along a common $z$ axis). The form of the hopping part of the Hamiltonian in Eq.~(\ref{eqn:tightbind}), is that of a matrix which is block diagonal in spin-space, where each block has, e.g., for $d$-states, dimension $5 \times 5$.  
Furthermore, the interaction term $\hat{\mathcal{H}}_{\mathrm{int}}$ includes the Coulomb and spin-orbit interactions, where the former interaction is responsible for spin-pairing and the possibility of forming a finite magnetic moment on each lattice site. Within a mean-field approximation, these interactions could in principle be included as a spin-dependent hopping term, but {we choose} to keep the separation between the spin-independent hopping in $\hat{\mathcal{H}}_0$ and the interaction term $\hat{\mathcal{H}}_{\mathrm{int}}$, such that $\hat{\mathcal{H}}_0$ is independent of the magnetic state. {Within this formalism, a spin-dependent hopping contribution could in principle still be included in $\hat{\mathcal{H}}_{\mathrm{int}}$ and in the corresponding self-energy $\Sigma$.}

\subsection{Exchange from constraining field\label{sec:Exchange from constraining field}}

To obtain the exchange tensor $J_{ij}^{\alpha\beta}$, we have to
calculate the derivative of the constraining field. Our starting point
is to calculate the change of the magnetic moment component $M_{j\beta}$
under a change of the prescribed directions $\{\mathbf{e}_{i}\}$ \citep{Bruno2003},
from which we obtain the set of equations 
\begin{equation}
\frac{\partial M_{j\beta}}{\partial e_{i\alpha}}=\frac{\partial\left(M_{j}e_{j\beta}\right)}{\partial e_{i\alpha}}\approx M_{j}\frac{\partial e_{j\beta}}{\partial e_{i\alpha}}=M_{i}\delta_{\alpha\beta}\delta_{ij},\label{eq:moment derivative}
\end{equation}
where we assume a constant magnetic moment length $M_{j}$. This approximation
is valid in the magnetic ground state \citep{Liechtenstein1987}. {Keeping the moment length fixed introduces to the constraining field a spurious contribution that is parallel to the moment direction, which has, however, no relevance for the spin dynamics, Eq.~(\ref{eq:EOM}), and can be projected out (see Appendix \ref{sec:projection}).}

The derivative of $M_{j\beta}$ can be obtained by expressing the
expectation value via Matsubara Green's functions,

\begin{align}
M_{j\beta} & =\frac{\hbar\gamma}{2}\sum_{\o}\sum_{\sigma\sigma'}\sigma_{\sigma\sigma'}^{\beta}\int_{\omega}\left[\mathcal{G}(i\omega)\right]_{j\o\sigma',j\o\sigma},\\
 & =\frac{\hbar\gamma}{2}\int_{\omega}\text{Tr}\left\{ \sigma_{j}^{\beta}\mathcal{G}(i\omega)\right\} ,
\end{align}
with matrix elements
\begin{equation}
\left[\sigma_{k}^{\beta}\right]_{i\o\sigma,j\o'\sigma'}=\sigma_{\sigma\sigma'}^{\beta}\delta_{ik}\delta_{jk}\delta_{\o\o'},
\end{equation}
where $\boldsymbol{\sigma}_{\sigma\sigma'}$ is the Pauli matrix vector. We use the following short-hand notation for the Matsubara
sums,

\begin{equation}
\int_{\omega}\equiv k_{B}T\sum_{i\omega}e^{i\omega0^{+}},
\end{equation}
with Boltzmann constant $k_{B}$ and temperature $T$. We include
the required convergence factor $e^{i\omega0^{+}}$ for the correct time-ordering of operators. 

From the inverse matrix derivative rule,
\begin{equation}
\partial\mathcal{G}=-\mathcal{G}\left(\partial\mathcal{G}^{-1}\right)\mathcal{G},
\end{equation}
together with the Dyson equation,
\begin{equation}
\mathcal{G}^{-1}=\mathcal{G}_{0}^{-1}-\Sigma,
\end{equation}
we obtain
\begin{align}
\frac{\partial M_{j\beta}}{\partial e_{i\alpha}} & =\frac{\hbar\gamma}{2}\int_{\omega}\text{Tr}\left\{ \sigma_{j}^{\beta}\mathcal{G}(i\omega)\left(\frac{\partial\mathcal{H}_{\text{con}}}{\partial e_{i\alpha}}+\frac{\partial\Sigma(i\omega)}{\partial e_{i\alpha}}\right)\mathcal{G}(i\omega)\right\},
\end{align}
where we assume a constant chemical potential $\mu$ (see Appendix
\ref{sec:matsubara} for the definition of the Green's functions).
Here we have used that the non-interacting Green's function $\mathcal{G}_{0}$
only depends on the moment directions via the constraining field contribution (which we include in $\mathcal{G}_0$),
\begin{equation}
\frac{\partial\mathcal{G}_{0}^{-1}}{\partial e_{i\alpha}}=-\frac{\partial\mathcal{H}_{\text{con}}}{\partial e_{i\alpha}}.
\end{equation}
All other contributions that depend on the moment directions are by
definition included in the self-energy $\Sigma$, which takes correlation effects from $\hat{\mathcal{H}}_{\mathrm{int}}$ into account. From the matrix of the
constraining part of the Hamiltonian,
\begin{equation}
\mathcal{H}_{\text{con}}=-\frac{\hbar\gamma}{2}\sum_{k\nu}\sigma_{k}^{\nu}B_{k\nu}^{\text{con}},    
\end{equation}
we obtain the corresponding derivative
\begin{equation}
\frac{\partial\mathcal{H}_{\text{con}}}{\partial e_{i\alpha}}=-\frac{\hbar\gamma}{2}\sum_{k\nu}\sigma_{k}^{\nu}\frac{\partial B_{k\nu}^{\text{con}}}{\partial e_{i\alpha}}.
\end{equation}
We can now write
\begin{align}
\frac{\partial M_{j\beta}}{\partial e_{i\alpha}} & =K_{ji}^{\beta\alpha}+\sum_{k\nu}X_{jk}^{\beta\nu}\frac{\partial B_{k\nu}^{\text{con}}}{\partial e_{i\alpha}},\label{eq:moment derivative GF}
\end{align}
with
\begin{align}
K_{ij}^{\alpha\beta} & =\frac{\hbar\gamma}{2}\int_{\omega}\text{Tr}\left\{ \sigma_{i}^{\alpha}\mathcal{G}(i\omega)\left(\frac{\partial\Sigma(i\omega)}{\partial e_{j\beta}}\right)\mathcal{G}(i\omega)\right\} ,\\
X_{ij}^{\alpha\beta} & =-\frac{\hbar^{2}\gamma^{2}}{4}\int_{\omega}\text{Tr}\left\{ \sigma_{i}^{\alpha}\mathcal{G}(i\omega)\sigma_{j}^{\beta}\mathcal{G}(i\omega)\right\} .
\end{align}
Inserting Eq.~(\ref{eq:moment derivative GF}) into Eq.~(\ref{eq:moment derivative}) gives
\begin{equation}
\sum_{k\nu}X_{jk}^{\beta\nu}\frac{\partial B_{k\nu}^{\text{con}}}{\partial e_{i\alpha}}=M_{i}\delta_{\alpha\beta}\delta_{ij}-K_{ji}^{\beta\alpha}.
\end{equation}
By multiplying with $X^{-1}$, we finally obtain
\begin{equation}
J_{ij}^{\alpha\beta}=-M_{i}\sum_{k\nu}\left[X^{-1}\right]_{ik}^{\alpha\nu}\left(M_{j}\delta_{\beta\nu}\delta_{jk}-K_{kj}^{\nu\beta}\right).\label{eq:J}
\end{equation}

If the derivative $\partial\Sigma/\partial e_{i\alpha}$ is not easily
accessible but $\partial\Sigma/\partial B_{i\alpha}^{\text{con}}$
is, then we can take a slightly different approach. We write

\begin{equation}
\frac{\partial M_{j\beta}}{\partial e_{i\alpha}}=\sum_{k\nu}\tilde{X}_{jk}^{\beta\nu}\frac{\partial B_{k\nu}^{\text{con}}}{\partial e_{i\alpha}},
\end{equation}
with
\begin{align}
\tilde{X}_{ij}^{\alpha\beta} & =\nonumber \\
- & \frac{\hbar^{2}\gamma^{2}}{4}\int_{\omega}\text{Tr}\left\{ \sigma_{i}^{\alpha}\mathcal{G}(i\omega)\left[\sigma_{j}^{\beta}-\frac{2}{\hbar\gamma}\frac{\partial\Sigma(i\omega)}{\partial B_{j\beta}^{\text{con}}}\right]\mathcal{G}(i\omega)\right\} .
\end{align}
We obtain then the alternative but equivalent result
\begin{equation}
J_{ij}^{\alpha\beta}=-M_{i}M_{j}\left[\tilde{X}^{-1}\right]_{ij}^{\alpha\beta}.
\end{equation}
{This reformulation is useful when the self-energy is obtained from a diagrammatic expansion of the self-energy in terms of non-interacting Green's functions, where the derivative of the self-energy with respect to the constraining field can be performed analytically.}

\subsection{DFT-like correction term}

In DFT calculations with constraining fields the effective magnetic field is given by the energy gradient which is in this case not identical to the negative of the constraining field \citep{Streib2020},
\begin{equation}
\mathbf{B}_{i}^{\text{eff}}=-\frac{1}{M_i}\boldsymbol{\nabla}_{\mathbf{e}_{i}}E=-\mathbf{B}_{i}^{\text{con}}-\frac{1}{M_{i}}\left\langle \boldsymbol{\nabla}_{\mathbf{e}_{i}}^{*}\hat{\mathcal{H}}_{\text{KS}}\right\rangle,\label{eq:field DFT}
\end{equation}
where $\hat{\mathcal{H}}_{\text{KS}}$ is the auxiliary Kohn-Sham
Hamiltonian \citep{Kohn1965} and $\boldsymbol{\nabla}_{\mathbf{e}_{i}}^{*}$
denotes the derivative with constant electron densities and moment
lengths \citep{Liechtenstein1987}.
Although in the present investigation we are considering tight-binding models, we may need to take
this correction term into account if the tight-binding model has been
fitted to Kohn-Sham band structures and the DFT formalism has to be applied for consistency.
The self-consistent exchange constants are then obtained from a derivative of Eq.~(\ref{eq:field DFT}),
\begin{equation}
J_{\text{sc},ij}^{\alpha\beta}=-M_{i}\frac{\partial B_{i\alpha}^{\text{con}}}{\partial e_{j\beta}}-\frac{\partial}{\partial e_{j\beta}}\left\langle \frac{\partial^{*}}{\partial e_{i\alpha}}\hat{\mathcal{H}}_{\text{KS}}\right\rangle.\label{eq:J_DFT definition}
\end{equation}
We assume now that we have a tight-binding Hamiltonian that reproduces
the band structure of $\hat{\mathcal{H}}_{\text{KS}}$, where the exchange
splitting is parameterized via the following Stoner term \citep{Brooks1983,Aute2006,Schena2010}, 
\begin{equation}
\hat{\mathcal{H}}_{\text{St}}=\sum_{i\o\o'}\frac{I_{\o\o'}}{\hbar\mu_{B}}M_{i\o}\mathbf{e}_{i}\cdot\hat{\mathbf{S}}_{i\o'},\label{eq:Stoner model}
\end{equation}
where $M_{i\o}$ is the magnetic moment length and $\hat{\mathbf{S}}_{i\o}$ the spin operator associated with the orbital $\o$ at site $i$ with Stoner parameter $I_{\o\o'}$. From
\begin{equation}
\left\langle \frac{\partial^{*}}{\partial e_{i\alpha}}\hat{\mathcal{H}}_{\text{St}}\right\rangle =\sum_{\o\o'}\frac{I_{\o\o'}M_{i\o}}{\hbar\gamma\mu_{B}}M_{i\o'\alpha},
\end{equation}
we obtain
\begin{equation}
\frac{\partial}{\partial e_{j\beta}}\left\langle \frac{\partial^{*}}{\partial e_{i\alpha}}\hat{\mathcal{H}}_{\text{St}}\right\rangle =\sum_{\o\o'}\frac{I_{\o\o'}M_{i\o}}{\hbar\gamma\mu_{B}}\frac{\partial M_{i\o'\alpha}}{\partial e_{j\beta}},\label{eq:Stoner double derivative}
\end{equation}
where we assume that both $I_{\o\o'}$ and $M_{i\o}$ are constant. Next, analogous to Eq.~(\ref{eq:moment derivative GF}), 
we derive an expression for the derivative of the orbital resolved magnetic moments,
\begin{equation}
\frac{\partial M_{j\o\beta}}{\partial e_{i\alpha}}=K_{ji}^{\beta\alpha}(\o)+\sum_{k\nu}X_{jk}^{\beta\nu}(\o)\frac{\partial B_{k\nu}^{\text{con}}}{\partial e_{i\alpha}},\label{eq:moment derivative orbital}
\end{equation}
where $(\o)$ denotes that we restrict the trace in the calculation
of the matrices to a single orbital with index $\o$. Combining Eqs.~(\ref{eq:moment derivative orbital}), (\ref{eq:Stoner double derivative}), and (\ref{eq:J_DFT definition}), we obtain
\begin{align}
J_{\text{sc},ij}^{\alpha\beta} & =J_{ij}^{\alpha\beta}+J_{0,ij}^{\alpha\beta}+\sum_{k\nu}\frac{1}{M_{k}}K_{ki}^{\nu\alpha}J_{kj}^{\nu\beta},\label{eq:J_DFT}
\end{align}
where 
\begin{align}
J_{0,ij}^{\alpha\beta} & =-\int_{\omega}\text{Tr}\left\{ \frac{\partial\Sigma}{\partial e_{i\alpha}}\mathcal{G}(i\omega)\frac{\partial\Sigma}{\partial e_{j\beta}}\mathcal{G}(i\omega)\right\} \label{eq:J0}
\end{align}
is the contribution that we would get without the constraining
field and $J_{ij}^{\alpha\beta}$ is the pure constraining field contribution derived in Sec.~\ref{sec:Exchange from constraining field}.
Here, the self-energy is given by the Stoner term,
\begin{equation}
\Sigma=\mathcal{H}_{\text{St}},
\end{equation}
which is by definition included in the self-energy due to the dependence of $\mathcal{H}_{\text{St}}$
on the moment directions.

The structure of Eq.~(\ref{eq:J_DFT}) corresponds to the DFT results
by Bruno \citep{Bruno2003} for {isotropic} Heisenberg exchange parameters $J_{ij}=J_{ij}^{xx}=J_{ij}^{yy}$
and an equation equivalent to Eq.~(\ref{eq:J0}) has been derived
previously by Katsnelson and Lichtenstein \citep{Katsnelson2000},
again only for Heisenberg parameters $J_{ij}$ (see also Ref.~\citep{Nomoto2020}). Exchange parameters calculated
without constraining fields have been shown to
give the exact spin-wave energies in the long-wavelength limit \citep{Bruno2003,Antropov2003,Katsnelson2004},
i.e., for the calculation of the exchange stiffness constant it is
not necessary to consider constraining fields.

\section{Numerical results\label{sec:numerics}}

We have three different equations available for the calculation of
exchange parameters: Eq.~(\ref{eq:J}) which is based on the constraining
field, Eq.~(\ref{eq:J_DFT}) which is valid for DFT calculations, or parametrized calculations mimicking DFT results (as employed here) that include 
the constraining fields, and Eq.~(\ref{eq:J0}) which is obtained
without constraining fields. We have implemented these three equations within the CAHMD package \citep{CAHMD} and applied them to
a mean-field tight-binding model for iron with a ferromagnetic ground state. We compare the results of each approach for iron dimers
and iron chains with a lattice constant of $2.86\;\textup{\AA}$. The Slater-Koster parameters \citep{Slater1954} of the
tight-binding model are taken from Ref.~\citep{Thonig2014}, we do not include spin-orbit coupling, and we
use the Stoner term (\ref{eq:Stoner model}), see Ref.~\citep{Streib2020}
for details. For the numerical evaluation of the exchange constants, we use a finite temperature parameter, e.g., $T=1\;K$, to avoid divergences in the derivative of the Fermi function, see Appendix~\ref{sec:matsubara}. Since we do not take finite temperature effects on the electronic structure and lattice vibrations \citep{Mankovsky2020b} into account, we will consider only the zero-temperature limit.

\subsection{Fe dimer\label{sec:dimer}}

\begin{figure}
\begin{centering}
\includegraphics{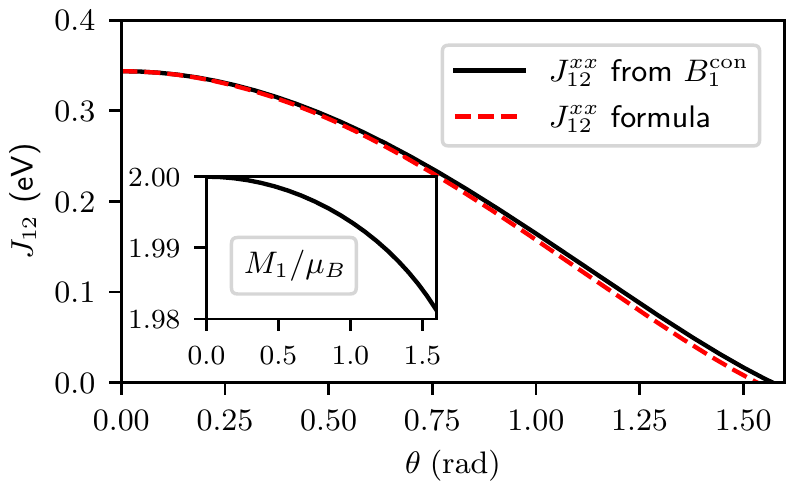}
\par\end{centering}
\caption{Comparison of the exchange parameter $J_{12}^{xx}$ obtained
by numerical differentiation of the constraining field, Eq.~(\ref{eq:J constraining field}), and the one
obtained from Eq.~(\ref{eq:J}) as a function of the angle $\theta$
between the two magnetic moments of an iron dimer. The inset shows the magnetic moment length $M_1$($=M_2$). \label{fig:J_B}}
\end{figure}

We first compare in Fig.~\ref{fig:J_B} for an iron dimer the calculated
exchange constant $J_{12}^{xx}$ from Eq.~(\ref{eq:J}) with the result obtained by
numerical differentiation of the constraining field,
\begin{equation}
J_{12}^{xx}=-M_{1}\frac{\partial B_{1x}^{\text{con}}}{\partial e_{2x}}.\label{eq:J constraining field}
\end{equation}
We keep the first moment aligned along the $z$ axis and rotate the
second moment by an angle $\theta$ in the $xz$ plane. The rotation is performed by adjusting the moment direction in the Stoner term (\ref{eq:Stoner model}) and applying the required constraining field to stabilize the configuration. In the limit
$\theta\to0$ (the ferromagnetic ground state), the agreement is exact, while for $\theta>0$ there
is a small difference which is due to the dependencies of the magnetic moments and the chemical potential
on the spin configuration that are both not taken into account in our derivation
of the exchange constants. The magnetic moments, with $M_1=M_2$, vary by about $1\%$ and the chemical potential by about $2\%$ in the range $\theta=0$ to $\pi/2$, see Fig.~\ref{fig:J_B} and Fig.~\ref{fig:mu} in Appendix \ref{sec:matsubara}.

\begin{figure}
\begin{centering}
\includegraphics{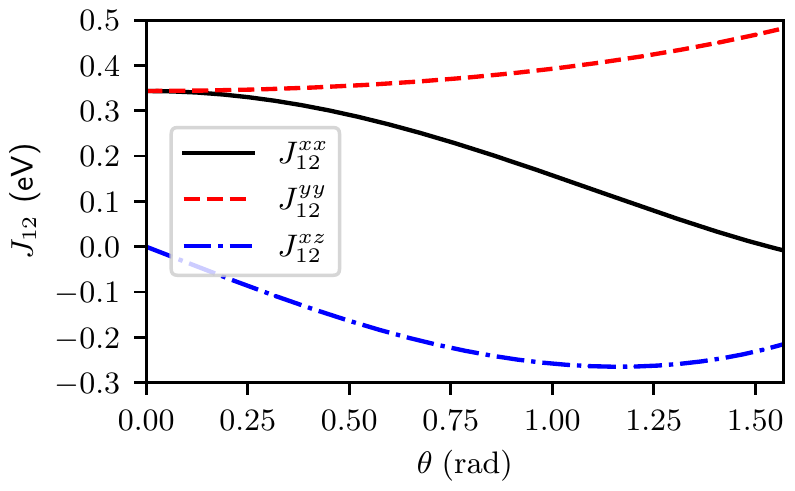}
\par\end{centering}
\caption{Components of the exchange tensor $J_{12}^{\alpha\beta}$ obtained
from Eq.~(\ref{eq:J}) as a function of the angle $\theta$ between
the two magnetic moments of an iron dimer.\label{fig:J}}
\end{figure}

Next, we consider the exchange tensor $J_{12}^{\alpha\beta}$ as a
function of the angle $\theta$ between the two moments. {After projection to perpendicular fields according to Eqs.~(\ref{eq:projected J 1}) and (\ref{eq:projected J 2}) in Appendix \ref{sec:projection}, which removes the spurious contribution mentioned after Eq.~(\ref{eq:moment derivative}),} only the
components $xx$, $yy$, and $xz$ are finite and are shown in Fig.~\ref{fig:J}.
In the ferromagnetic ground state, $\theta=0$, we have $J_{xx}=J_{yy}$
and $J_{xz}=0$, indicating a Heisenberg-like local spin Hamiltonian.
However, for $\theta>0$, $J_{xx}\neq J_{yy}$ and $J_{xz}\neq0$,
implying a non-Heisenberg-like behavior, which requires the inclusion
of the linear term $C_{i\alpha}$ in the local spin Hamiltonian.

\begin{figure}
\begin{centering}
\includegraphics{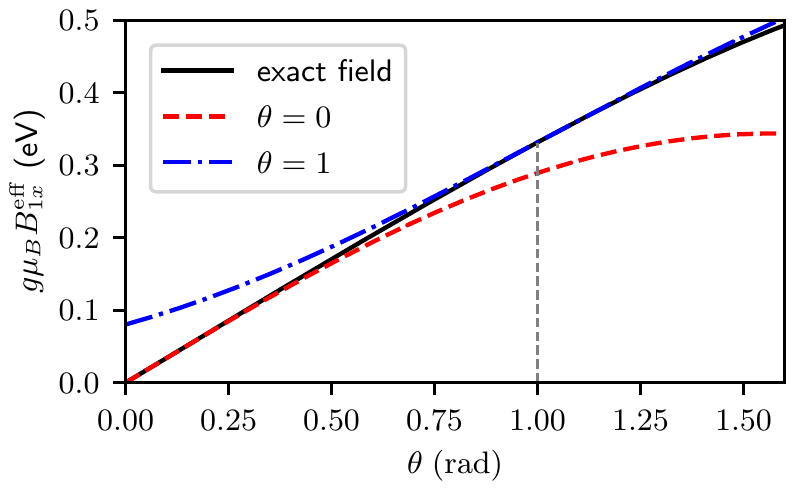}
\par\end{centering}
\caption{Comparison of the effective field obtained from the constraining field with the field from
the local spin Hamiltonian, Eq.~(\ref{eq:local Hamiltonian}), with
parameters calculated with Eq.~(\ref{eq:J}) for $\theta=0$ and $\theta=1$, as a function
of the angle $\theta$ between the two magnetic moments of an iron dimer.\label{fig:B}}
\end{figure}

We calculate the exchange parameters from Eq.~(\ref{eq:J}) for two specific angles
($\theta=0$ and $\theta=1$) {as examples} and apply them to the local spin Hamiltonian, Eq.~(\ref{eq:local Hamiltonian}). In Fig.~\ref{fig:B} we compare the effective field for these two cases with the effective
field given by the negative of the constraining field. For $\theta=0$ the local spin Hamiltonian
is a simple Heisenberg model with a single exchange parameter $J=J^{xx}=J^{yy}$,
while for $\theta=1$ the full exchange tensor $J_{ij}^{\alpha\beta}$
and the linear term $C_{i\alpha}$ have to be taken into account with $C_{1x}=0.3315\;\mathrm{eV}$. The linear term includes a contribution from the effective field as defined in Eq.~(\ref{eq:linear term}).
In both cases, the local spin Hamiltonian correctly describes small
deviations around the reference spin configurations, $\theta=0$ and
$\theta=1$. The deviations of the exact field from the Heisenberg
model confirms that beyond-Heisenberg contributions are present in
the underlying system, which is consistent with the configuration-dependent exchange
parameters in Fig.~\ref{fig:J}.

We compare for the ferromagnetic ground state ($\theta=0$) the Heisenberg exchange constants from Eqs.~(\ref{eq:J}), (\ref{eq:J0}),
and (\ref{eq:J_DFT}),
\begin{align}
J & =0.3437\;\text{eV},\\
J_{0} & =0.2575\;\text{eV},\\
J_{\text{sc}} & =0.3485\;\text{eV}.
\end{align}
While the correction term changes $J_{\text{sc}}$ in comparison to $J$ only by a small amount,
the difference to $J_{0}$, which is obtained without the inclusion
of constraining fields, is more significant. For comparison, a recent tight-binding calculation with constraining fields obtained $J=0.616\;\mathrm{eV}$ for an iron dimer with a lattice constant of $2\;\textup{\AA}$ \citep{Cardias2021}, where the smaller lattice constant causes a stronger exchange coupling than in our case with lattice constant $2.86\;\textup{\AA}$. In DFT calculations, similar deviations of the nearest-neighbor exchange with and without constraining fields have been observed for bulk bcc Fe and fcc Ni, while the energies of long-wavelength spin waves are unaffected by constraining fields \citep{Jacobsson2017,Solovyev2020}.

Although the difference between $J$ and $J_{\text{sc}}$ is very
small, the difference between the effective fields with and without
the DFT-like correction term can become significant in non-collinear states
\citep{Streib2020}. This can be understood by considering that the
exchange parameters give the derivative of the effective field and
a small difference in the derivative can change the effective field
significantly for strongly non-collinear states.

{While the exchange constants $J_{0,ij}^{\alpha\beta}$ and $J_{\mathrm{sc},ij}^{\alpha\beta}$ are always symmetric with respect to the interchange $i\alpha\leftrightarrow j\beta$, we find a small asymmetry for $J_{ij}^{\alpha\beta}$ in non-collinear states ($\theta\neq 0$) in our numerical calculations, which we discuss in Appendix \ref{sec:symmetry}.}

\subsection{Fe chain\label{sec:chain}}

\begin{figure}
\begin{centering}
\includegraphics{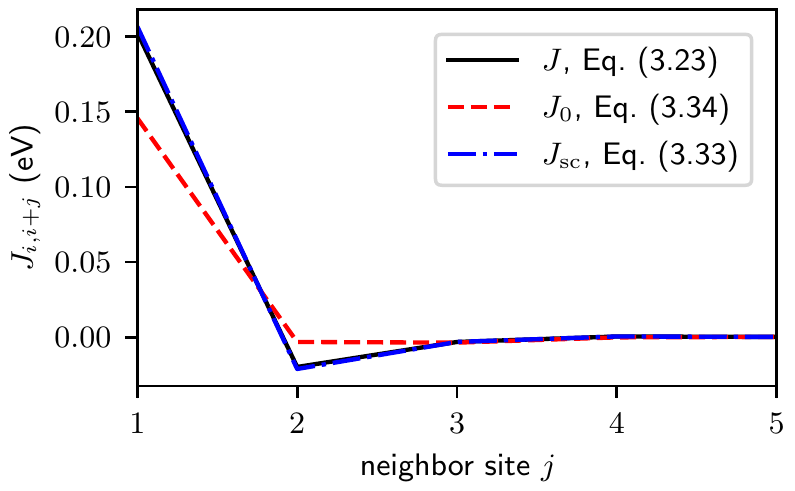}
\par\end{centering}
\caption{Exchange parameters $J_{i,i+j}$ between sites $i$ and $i+j$ obtained from Eqs.~(\ref{eq:J}), (\ref{eq:J0}), and (\ref{eq:J_DFT}) for an iron chain of {50} sites  in the ferromagnetic ground-state configuration with periodic boundary conditions. \label{fig:chain}}
\end{figure}

\begin{figure}
\begin{centering}
\includegraphics{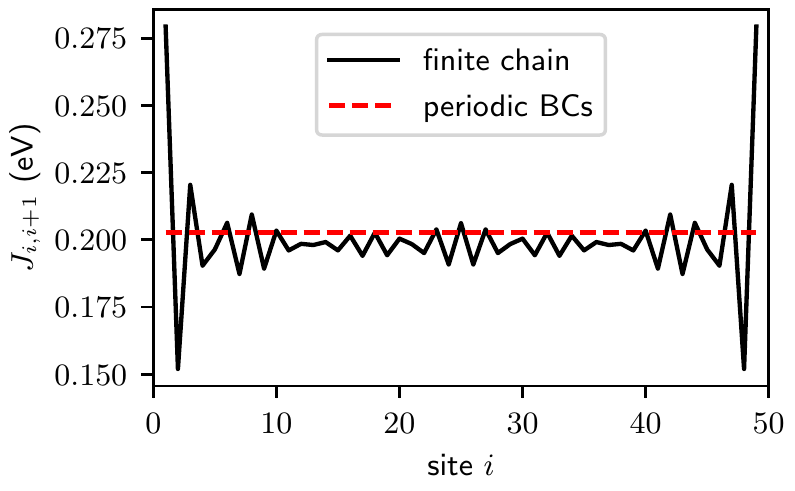}
\par\end{centering}
\caption{Nearest-neighbor exchange parameters $J_{i,i+1}$ obtained from Eq.~(\ref{eq:J}) for a finite iron chain of 50 sites in the ferromagnetic ground-state configuration in comparison to a chain with periodic boundary conditions. \label{fig:chain_bc}}
\end{figure}

In Fig.~\ref{fig:chain}, we show the Heisenberg exchange parameters $J_{ij}$ for an iron chain in its ferromagnetic ground state, where again the formulas based on the constraining field and the parameterized DFT-like formalism give similar results, Eqs.~(\ref{eq:J}) and (\ref{eq:J_DFT}), while the results without constraining field, Eq.~(\ref{eq:J0}), differ significantly. In Fig.~\ref{fig:chain_bc}, we compare results for the nearest neighbor exchange $J$, Eq.~(\ref{eq:J}), in a finite iron chain of $50$ spins with and without periodic boundary conditions. As expected, in the case with periodic boundary conditions the nearest neighbor exchange is completely uniform for all sites, while for the case without periodic boundary conditions there are strong variations near the boundaries of the chain and the deviations become smaller near the center. This reflects Friedel oscillations in the magnetic profile, induced by the abrupt change of the effective potential and hopping parameter at edges. Such variations of the exchange constants near the boundary of a magnet could be important for a proper description of surface and edge spin waves and topological magnons in two- and three-dimensional magnets \citep{Malki2020}.

\section{Summary\label{sec:summary}}

We have derived the mapping of tight-binding electronic structure theory to local spin
Hamiltonians. We show that in order to capture effects beyond bilinear Heisenberg exchange, the inclusion of a linear term to the spin Hamiltonian improves the accuracy of calculating, e.g., a local Weiss field. Linear contributions are usually not considered due to arguments based on the energy of time-reversed  states. We argue here that this is not a problem for 
local spin Hamiltonians, which are designed to describe energetics of spin fluctuations
around a given spin configuration, in particular by consideration of configuration-dependent parameters. Local spin Hamiltonians
are shown to be useful for the calculation of spin-wave spectra and spin
dynamics simulations near to the ground-state configuration.

We also provide
explicit formulas for the exchange constants based on a derivation
from the effective field for tight-binding models with and without
constraining fields. If we consider the effective field that is required
for DFT calculations \citep{Streib2020}, we recover previous results
\citep{Katsnelson2000,Bruno2003}, demonstrating the consistency of
our approach. 
We apply the derived formulas to a tight-binding model for iron dimers and chains,
and find good agreement with the exchange constant
derived by numerical differentiation of the constraining field. The {numerical} tight-binding electronic structure theory calculations {in Sec.~\ref{sec:numerics}} are based on a formulation where spin-functions are defined along a global quantization axis. In this formulation, the tight binding parameters (that typically are parametrized to reproduce static electronic structures obtained from \textit{ab initio} theory) are fixed and independent of magnetic configuration. This implies that the configuration dependence of the kinetic energy does not enter the equations of exchange interactions or the local Weiss field. A description that relies on a local quantization axis, for which spin-functions are defined on each atomic site, would release this constraint and represents an obvious extension of this work.   

The
local spin Hamiltonian, together with the exchange constant formulas,
is demonstrated to correctly describe the effective field near a given spin
configuration. We find, however, that for larger deviations from a given spin-configuration where the exchange parameters were calculated from, there can be a pronounced configuration dependence of the parameters. This fact implies that beyond-Heisenberg contributions
are required and are effectively taken into account.

While for the description of arbitrary spin configurations a global spin Hamiltonian is required,
we expect that the local approach described here will
find applications to characterize spin waves and spin fluctuations for
magnets with non-collinear ground states. A consistent extension of
the LKAG approach to these non-collinear states would not be possible
without the linear term in the Hamiltonian or by inclusion of higher-order terms \citep{Szilva2013}.
The exchange constant formulas that we have derived for tight-binding
models will be useful both for model Hamiltonians and for \emph{ab
initio} electronic structure calculations.

\begin{acknowledgments}
We thank Pavel Bessarab, Ksenia Vodenkova, and Lars Nordstr\"om for insightful discussions. The authors acknowledge financial support from the Knut and Alice Wallenberg Foundation through grant no. 2018.0060. O.E. also acknowledges support of eSSENCE, the Swedish Research Council (VR), the Foundation for Strategic Research (SSF) and ERC synergy grant (854843-FASTCORR).  D.T. acknowledges support from the Swedish Research Council (VR)  through  Grant  No.   2019-03666. A.D. acknowledges support from the Swedish Research Council (VR) through grant numbers VR 2015-04608, VR 2016-05980 and VR 2019-05304. The work of M.I.K. is supported by ERC synergy grant (854843-FASTCORR).
The computations were enabled by resources provided by the
Swedish National Infrastructure for Computing (SNIC) at Chalmers Center for Computational Science and Engineering (C3SE), High Performance Computing Center North (HPCN), and the National Supercomputer Center (NSC) partially funded by the Swedish Research Council through grant
agreement no. 2016-07213.
\end{acknowledgments}

\appendix

\section{Projection to perpendicular fields\label{sec:projection}}

If we consider a spin Hamiltonian
\begin{equation}
\mathcal{H}_{s}=-\sum_{i\alpha}\tilde{C}_{i\alpha}e_{i\alpha}-\frac{1}{2}\sum_{ij}\sum_{\alpha\beta}\tilde{J}_{ij}^{\alpha\beta}e_{i\alpha}e_{j\beta},
\end{equation}
that does not necessarily result in effective fields that are purely perpendicular
to the magnetic moment directions,
\begin{align}
B_{i\alpha}^{\text{eff}} & =\frac{1}{M_{i}}\tilde{C}_{i\alpha}+\frac{1}{M_{i}}\sum_{k\nu}\tilde{J}_{ik}^{\alpha\nu}e_{k\nu},
\end{align}
then we can project out the parallel component,
\begin{equation}
\mathbf{B}_{i\perp}^{\text{eff}}=\mathbf{B}_{i}^{\text{eff}}-\mathbf{e}_{i}\left(\mathbf{B}_{i}^{\text{eff}}\cdot\mathbf{e}_{i}\right).
\end{equation}
From this projection, we obtain parameters that produce purely
perpendicular fields,
\begin{align}
J_{ij}^{\alpha\beta} & =\tilde{J}_{ij}^{\alpha\beta}-\sum_{\nu}\tilde{J}_{ij}^{\nu\beta}e_{i\nu}e_{i\alpha},\\
\mathbf{C}_{i} & =\tilde{\mathbf{C}}_{i}-\mathbf{e}_{i}\left(\tilde{\mathbf{C}}_{i}\cdot\mathbf{e}_{i}\right).
\end{align}
For a state with $\mathbf{e}_{i}=\hat{\mathbf{e}}_{z}$,
we have{
\begin{align}
J_{ij}^{\alpha\beta} & =\tilde{J}_{ij}^{\alpha\beta}\;(\text{for}\;\alpha\neq z),\label{eq:projected J 1}\\
J_{ij}^{z\beta} & =0,\label{eq:projected J 2}\\
C_{iz} & =0.
\end{align}
This projection to perpendicular fields can break the symmetry $J_{ij}^{\alpha\beta}=J_{ji}^{\beta\alpha}$ and is not required for practical calculations since parallel components do not contribute to the equation of motion (\ref{eq:EOM}). }

\section{Matsubara sums\label{sec:matsubara}}

\begin{figure}
\begin{centering}
\includegraphics{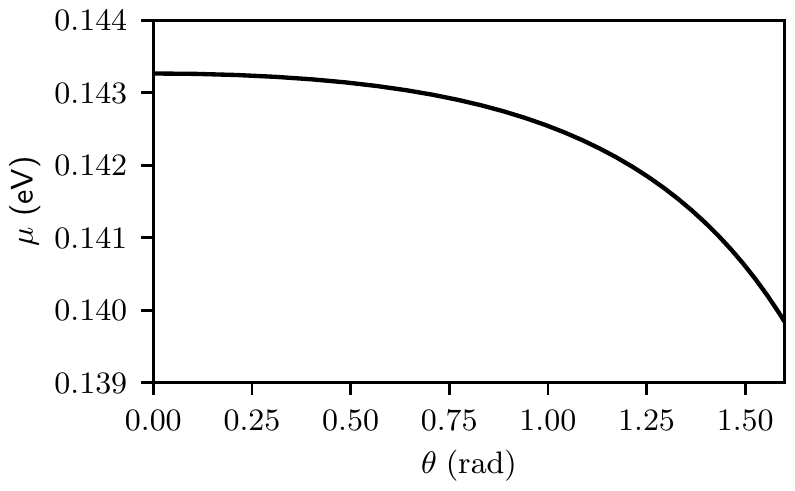}
\par\end{centering}
\caption{Dependence of the chemical potential $\mu$ of an iron dimer on the angle $\theta$ between the two magnetic moments.\label{fig:mu}}
\end{figure}
\begin{figure}
\begin{centering}
\includegraphics{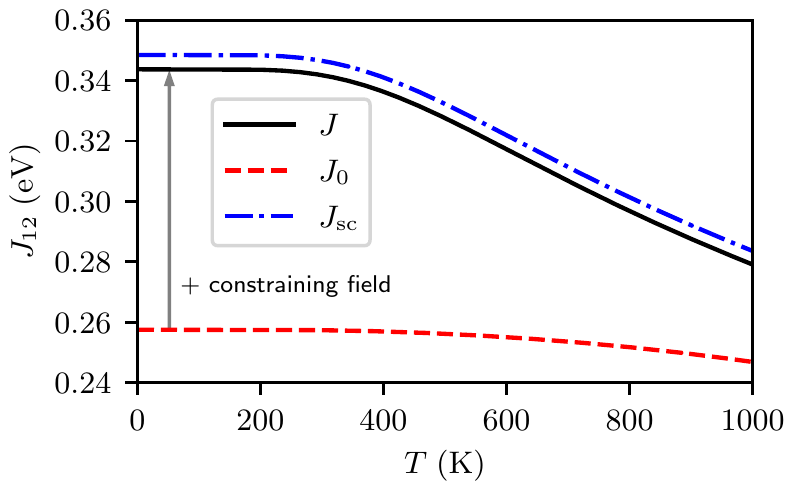}
\par\end{centering}
\caption{Dependence of the Heisenberg exchange $J_{12}$ of a ferromagnetic iron dimer on the temperature parameter $T$ in the exchange formulas Eqs.~(\ref{eq:J}), (\ref{eq:J0}), and (\ref{eq:J_DFT}).\label{fig:J_temp}}
\end{figure}

For the evaluation of the exchange formulas we have to calculate Matsubara
sums over pairs of Green's functions, for example
\begin{equation}
X_{ij}^{\alpha\beta} =-\frac{\hbar^{2}\gamma^{2}}{4}\int_{\omega}\text{Tr}\left\{ \sigma_{i}^{\alpha}\mathcal{G}(i\omega)\sigma_{j}^{\beta}\mathcal{G}(i\omega)\right\} .
\end{equation}
We use the eigenbasis expansion of the Matsubara Green's function,
\begin{equation}
\mathcal{G}(i\omega)=\sum_{n}\frac{\ket{n}\bra{n}}{i\omega-\xi_{n}},
\end{equation}
where $\{\ket{n}\}$ are the single electron eigenstates of the Hamiltonian
with $\xi_{n}=\varepsilon_{n}-\mu$. Here, $\varepsilon_{n}$
is the energy of the state $\ket{n}$ and $\mu$ is the chemical potential
that controls the occupation of the states, which we plot in Fig.~\ref{fig:mu} for the iron dimer. It is now straight-forward
to perform the summation over Matsubara frequencies,
\begin{align}
\int_{\omega}\frac{1}{\left(i\omega-\xi_{n}\right)\left(i\omega-\xi_{n'}\right)}
=\begin{cases}
\frac{f(\xi_{n})-f(\xi_{n'})}{\xi_{n}-\xi_{n'}}, & \xi_{n}\neq\xi_{n'}\\
f'(\xi_{n}), & \xi_{n}=\xi_{n'}
\end{cases},
\end{align}
where the Fermi function and its derivative are given by
\begin{align}
f(\xi_{n}) & =\frac{1}{e^{\beta\xi_{n}}+1},\\
f'(\xi_{n}) & =\frac{-\beta e^{\beta\xi_{n}}}{\left(e^{\beta\xi_{n}}+1\right)^{2}}=\beta f(\xi_{n})\left[f(\xi_{n})-1\right],
\end{align}
with the inverse temperature $\beta=1/(k_{B}T)$.

The Matsubara formalism that we employ here introduces a temperature dependence. Since we consider only the zero-temperature limit, it is important to confirm the convergence of our calculations for $T\to 0$. This is demonstrated in Fig.~\ref{fig:J_temp} for the Heisenberg exchange of an iron dimer in the ferromagnetic ground state, which shows only a weak temperature dependence for $T<300\;\mathrm{K}$.

{\section{Symmetry of exchange constants}\label{sec:symmetry}}

\begin{figure}
\begin{centering}
\includegraphics{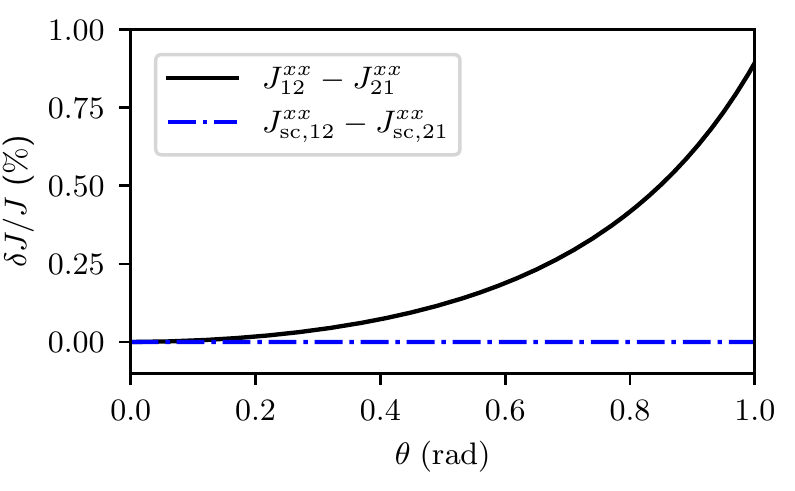}
\par\end{centering}
\caption{{Numerical check of the symmetry of the exchange constants $J_{ij}^{xx}$ and $J_{\mathrm{sc},ij}^{xx}$ of an iron dimer as a function of the angle $\theta$ between the two magnetic moments.}\label{fig:symmetry}}
\end{figure}

{Exchange constants that are derived from the curvature of the energy,
\begin{equation}
J_{ij}^{\alpha\beta}=-\frac{\partial^{2}\langle\hat{\mathcal{H}}\rangle}{\partial e_{i\alpha}\partial e_{j\beta}},
\end{equation}
are by definition symmetric with respect to the interchange $i\alpha\leftrightarrow j\beta$. While this fundamental symmetry is directly reflected in the exchange formula $J_{0,ij}^{\alpha\beta}$ in Eq.~(\ref{eq:J0}), it is not obvious that the derived formulas for $J_{ij}^{\alpha\beta}$ and $J_{\mathrm{sc},ij}^{\alpha\beta}$ in Eqs.~(\ref{eq:J}) and (\ref{eq:J_DFT}) fulfill this symmetry. We have therefore checked this symmetry numerically within our tight-binding calculations for an iron dimer. Our results in Fig.~\ref{fig:symmetry} show that $J_{\mathrm{sc},ij}^{\alpha\beta}$ fulfills the symmetry exactly, whereas $J_{ij}^{\alpha\beta}$ breaks it for $\theta\neq 0$, although the asymmetry reaches only the order of 1\%. This asymmetry is not related to the asymmetry that can be introduced by the projection discussed in Appendix \ref{sec:projection}, because we have not used this projection here.}

{From the constraining field theorem \citep{Streib2020}, we obtain the following relation for the exchange constant $J_{ij}^{\alpha\beta}$ obtained from the constraining field via Eq.~(\ref{eq:J from constrain}),
\begin{equation}
J_{ij}^{\alpha\beta}=-\frac{\partial^2 \langle \hat{\mathcal{H}}_\mathrm{tb}\rangle}{\partial e_{i\alpha}\partial e_{j\beta}}+\frac{\partial}{\partial e_{j\beta}}\left\langle \frac{\partial\hat{\mathcal{H}}_\mathrm{tb}}{\partial e_{i\alpha}}\right\rangle.\label{eq:constraining field theorem}
\end{equation}
The fundamental tight-binding Hamiltonian $\hat{\mathcal{H}}_\mathrm{tb}$ as defined in Eq.~(\ref{eq:H_tb}) is independent of the moment directions $\{\mathbf{e}_i\}$, which implies that $J_{ij}^{\alpha\beta}$ is symmetric since the second term on the right-hand side of Eq.~(\ref{eq:constraining field theorem}) vanishes in that case. However, our numerical calculations are based on a mean-field tight-binding model where the electron-electron interactions are effectively included within a mean-field approximation via the Stoner term, Eq.~(\ref{eq:Stoner model}). This Stoner term depends on the moment directions $\{\mathbf{e}_i\}$ and causes in our calculations the asymmetry of $J_{ij}^{\alpha\beta}$. The symmetry-breaking term is subtracted in the definition of $J_{\mathrm{sc},ij}^{\alpha\beta}$, Eq.~(\ref{eq:J_DFT definition}), such that
\begin{equation}
J_{\mathrm{sc},ij}^{\alpha\beta}=-\frac{\partial^2 \langle \hat{\mathcal{H}}_\mathrm{tb}\rangle}{\partial e_{i\alpha}\partial e_{j\beta}},
\end{equation}
where we assume constant charges and moment lengths, i.e., $\boldsymbol{\nabla}_{\mathbf{e}_{i}}\equiv \boldsymbol{\nabla}_{\mathbf{e}_{i}}^{*}$, which is consistent with the approximations made to derive the exchange formulas in Sec.~\ref{sec:exchange}. The quantity $\langle \hat{\mathcal{H}}_\mathrm{tb}\rangle$ describes the band energy of the tight-binding model, while the total energy contains additional constant energy contributions that arise from the mean-field decoupling and are not included in $\hat{\mathcal{H}}_\mathrm{tb}$ here. The exchange constant $J_{ij}^{\alpha\beta}$ is derived from the constraining field, which corresponds to the gradient of the total energy \citep{Streib2020}, explaining the difference between $J_{ij}^{\alpha\beta}$ and $J_{\mathrm{sc},ij}^{\alpha\beta}$. Without the mean-field approximation this difference would not arise.
}


\begin{thebibliography}{43}%
\makeatletter
\providecommand \@ifxundefined [1]{%
 \@ifx{#1\undefined}
}%
\providecommand \@ifnum [1]{%
 \ifnum #1\expandafter \@firstoftwo
 \else \expandafter \@secondoftwo
 \fi
}%
\providecommand \@ifx [1]{%
 \ifx #1\expandafter \@firstoftwo
 \else \expandafter \@secondoftwo
 \fi
}%
\providecommand \natexlab [1]{#1}%
\providecommand \enquote  [1]{``#1''}%
\providecommand \bibnamefont  [1]{#1}%
\providecommand \bibfnamefont [1]{#1}%
\providecommand \citenamefont [1]{#1}%
\providecommand \href@noop [0]{\@secondoftwo}%
\providecommand \href [0]{\begingroup \@sanitize@url \@href}%
\providecommand \@href[1]{\@@startlink{#1}\@@href}%
\providecommand \@@href[1]{\endgroup#1\@@endlink}%
\providecommand \@sanitize@url [0]{\catcode `\\12\catcode `\$12\catcode
  `\&12\catcode `\#12\catcode `\^12\catcode `\_12\catcode `\%12\relax}%
\providecommand \@@startlink[1]{}%
\providecommand \@@endlink[0]{}%
\providecommand \url  [0]{\begingroup\@sanitize@url \@url }%
\providecommand \@url [1]{\endgroup\@href {#1}{\urlprefix }}%
\providecommand \urlprefix  [0]{URL }%
\providecommand \Eprint [0]{\href }%
\providecommand \doibase [0]{https://doi.org/}%
\providecommand \selectlanguage [0]{\@gobble}%
\providecommand \bibinfo  [0]{\@secondoftwo}%
\providecommand \bibfield  [0]{\@secondoftwo}%
\providecommand \translation [1]{[#1]}%
\providecommand \BibitemOpen [0]{}%
\providecommand \bibitemStop [0]{}%
\providecommand \bibitemNoStop [0]{.\EOS\space}%
\providecommand \EOS [0]{\spacefactor3000\relax}%
\providecommand \BibitemShut  [1]{\csname bibitem#1\endcsname}%
\let\auto@bib@innerbib\@empty
\bibitem [{\citenamefont {Takahashi}(1977)}]{Takahashi1977}%
  \BibitemOpen
  \bibfield  {author} {\bibinfo {author} {\bibfnamefont {M.}~\bibnamefont
  {Takahashi}},\ }\bibfield  {title} {\bibinfo {title} {{Half-filled Hubbard
  model at low temperature}},\ }\href
  {https://doi.org/10.1088/0022-3719/10/8/031} {\bibfield  {journal} {\bibinfo
  {journal} {J. Phys. C: Solid State Phys.}\ }\textbf {\bibinfo {volume}
  {10}},\ \bibinfo {pages} {1289} (\bibinfo {year} {1977})}\BibitemShut
  {NoStop}%
\bibitem [{\citenamefont {MacDonald}\ \emph {et~al.}(1988)\citenamefont
  {MacDonald}, \citenamefont {Girvin},\ and\ \citenamefont
  {Yoshioka}}]{MacDonald1988}%
  \BibitemOpen
  \bibfield  {author} {\bibinfo {author} {\bibfnamefont {A.~H.}\ \bibnamefont
  {MacDonald}}, \bibinfo {author} {\bibfnamefont {S.~M.}\ \bibnamefont
  {Girvin}},\ and\ \bibinfo {author} {\bibfnamefont {D.}~\bibnamefont
  {Yoshioka}},\ }\bibfield  {title} {\bibinfo {title} {{$\frac{t}{U}$ expansion
  for the Hubbard model}},\ }\href {https://doi.org/10.1103/PhysRevB.37.9753}
  {\bibfield  {journal} {\bibinfo  {journal} {Phys. Rev. B}\ }\textbf {\bibinfo
  {volume} {37}},\ \bibinfo {pages} {9753} (\bibinfo {year}
  {1988})}\BibitemShut {NoStop}%
\bibitem [{\citenamefont {Hoffmann}\ and\ \citenamefont
  {Bl\"ugel}(2020)}]{Hoffmann2020}%
  \BibitemOpen
  \bibfield  {author} {\bibinfo {author} {\bibfnamefont {M.}~\bibnamefont
  {Hoffmann}}\ and\ \bibinfo {author} {\bibfnamefont {S.}~\bibnamefont
  {Bl\"ugel}},\ }\bibfield  {title} {\bibinfo {title} {Systematic derivation of
  realistic spin models for beyond-heisenberg solids},\ }\href
  {https://doi.org/10.1103/PhysRevB.101.024418} {\bibfield  {journal} {\bibinfo
   {journal} {Phys. Rev. B}\ }\textbf {\bibinfo {volume} {101}},\ \bibinfo
  {pages} {024418} (\bibinfo {year} {2020})}\BibitemShut {NoStop}%
\bibitem [{\citenamefont {K\"ubler}(2000)}]{Kuebler}%
  \BibitemOpen
  \bibfield  {author} {\bibinfo {author} {\bibfnamefont {J.}~\bibnamefont
  {K\"ubler}},\ }\href@noop {} {\emph {\bibinfo {title} {Theory of Itinerant
  Electron Magnetism}}}\ (\bibinfo  {publisher} {Oxford University Press},\
  \bibinfo {address} {Oxford},\ \bibinfo {year} {2000})\BibitemShut {NoStop}%
\bibitem [{\citenamefont {Drautz}\ and\ \citenamefont
  {F\"ahnle}(2004)}]{Drautz2004}%
  \BibitemOpen
  \bibfield  {author} {\bibinfo {author} {\bibfnamefont {R.}~\bibnamefont
  {Drautz}}\ and\ \bibinfo {author} {\bibfnamefont {M.}~\bibnamefont
  {F\"ahnle}},\ }\bibfield  {title} {\bibinfo {title} {Spin-cluster expansion:
  Parametrization of the general adiabatic magnetic energy surface with ab
  initio accuracy},\ }\href {https://doi.org/10.1103/PhysRevB.69.104404}
  {\bibfield  {journal} {\bibinfo  {journal} {Phys. Rev. B}\ }\textbf {\bibinfo
  {volume} {69}},\ \bibinfo {pages} {104404} (\bibinfo {year}
  {2004})}\BibitemShut {NoStop}%
\bibitem [{\citenamefont {Drautz}\ and\ \citenamefont
  {F\"ahnle}(2005)}]{Drautz2005}%
  \BibitemOpen
  \bibfield  {author} {\bibinfo {author} {\bibfnamefont {R.}~\bibnamefont
  {Drautz}}\ and\ \bibinfo {author} {\bibfnamefont {M.}~\bibnamefont
  {F\"ahnle}},\ }\bibfield  {title} {\bibinfo {title} {Parametrization of the
  magnetic energy at the atomic level},\ }\href
  {https://doi.org/10.1103/PhysRevB.72.212405} {\bibfield  {journal} {\bibinfo
  {journal} {Phys. Rev. B}\ }\textbf {\bibinfo {volume} {72}},\ \bibinfo
  {pages} {212405} (\bibinfo {year} {2005})}\BibitemShut {NoStop}%
\bibitem [{\citenamefont {Antal}\ \emph {et~al.}(2008)\citenamefont {Antal},
  \citenamefont {Lazarovits}, \citenamefont {Udvardi}, \citenamefont
  {Szunyogh}, \citenamefont {\'Ujfalussy},\ and\ \citenamefont
  {Weinberger}}]{Antal2008}%
  \BibitemOpen
  \bibfield  {author} {\bibinfo {author} {\bibfnamefont {A.}~\bibnamefont
  {Antal}}, \bibinfo {author} {\bibfnamefont {B.}~\bibnamefont {Lazarovits}},
  \bibinfo {author} {\bibfnamefont {L.}~\bibnamefont {Udvardi}}, \bibinfo
  {author} {\bibfnamefont {L.}~\bibnamefont {Szunyogh}}, \bibinfo {author}
  {\bibfnamefont {B.}~\bibnamefont {\'Ujfalussy}},\ and\ \bibinfo {author}
  {\bibfnamefont {P.}~\bibnamefont {Weinberger}},\ }\bibfield  {title}
  {\bibinfo {title} {{First-principles calculations of spin interactions and
  the magnetic ground states of Cr trimers on Au(111)}},\ }\href
  {https://doi.org/10.1103/PhysRevB.77.174429} {\bibfield  {journal} {\bibinfo
  {journal} {Phys. Rev. B}\ }\textbf {\bibinfo {volume} {77}},\ \bibinfo
  {pages} {174429} (\bibinfo {year} {2008})}\BibitemShut {NoStop}%
\bibitem [{\citenamefont {Grytsiuk}\ \emph {et~al.}(2020)\citenamefont
  {Grytsiuk}, \citenamefont {Hanke}, \citenamefont {Hoffmann}, \citenamefont
  {Bouaziz}, \citenamefont {Gomonay}, \citenamefont {Bihlmayer}, \citenamefont
  {Lounis}, \citenamefont {Mokrousov},\ and\ \citenamefont
  {Bl{\"u}gel}}]{Grytsiuk2020}%
  \BibitemOpen
  \bibfield  {author} {\bibinfo {author} {\bibfnamefont {S.}~\bibnamefont
  {Grytsiuk}}, \bibinfo {author} {\bibfnamefont {J.-P.}\ \bibnamefont {Hanke}},
  \bibinfo {author} {\bibfnamefont {M.}~\bibnamefont {Hoffmann}}, \bibinfo
  {author} {\bibfnamefont {J.}~\bibnamefont {Bouaziz}}, \bibinfo {author}
  {\bibfnamefont {O.}~\bibnamefont {Gomonay}}, \bibinfo {author} {\bibfnamefont
  {G.}~\bibnamefont {Bihlmayer}}, \bibinfo {author} {\bibfnamefont
  {S.}~\bibnamefont {Lounis}}, \bibinfo {author} {\bibfnamefont
  {Y.}~\bibnamefont {Mokrousov}},\ and\ \bibinfo {author} {\bibfnamefont
  {S.}~\bibnamefont {Bl{\"u}gel}},\ }\bibfield  {title} {\bibinfo {title}
  {Topological--chiral magnetic interactions driven by emergent orbital
  magnetism},\ }\href {https://doi.org/10.1038/s41467-019-14030-3} {\bibfield
  {journal} {\bibinfo  {journal} {Nat. Commun.}\ }\textbf {\bibinfo {volume}
  {11}},\ \bibinfo {pages} {511} (\bibinfo {year} {2020})}\BibitemShut
  {NoStop}%
\bibitem [{\citenamefont {Brinker}\ \emph {et~al.}(2020)\citenamefont
  {Brinker}, \citenamefont {dos Santos~Dias},\ and\ \citenamefont
  {Lounis}}]{Brinker2020}%
  \BibitemOpen
  \bibfield  {author} {\bibinfo {author} {\bibfnamefont {S.}~\bibnamefont
  {Brinker}}, \bibinfo {author} {\bibfnamefont {M.}~\bibnamefont {dos
  Santos~Dias}},\ and\ \bibinfo {author} {\bibfnamefont {S.}~\bibnamefont
  {Lounis}},\ }\bibfield  {title} {\bibinfo {title} {Prospecting chiral
  multisite interactions in prototypical magnetic systems},\ }\href
  {https://doi.org/10.1103/PhysRevResearch.2.033240} {\bibfield  {journal}
  {\bibinfo  {journal} {Phys. Rev. Research}\ }\textbf {\bibinfo {volume}
  {2}},\ \bibinfo {pages} {033240} (\bibinfo {year} {2020})}\BibitemShut
  {NoStop}%
\bibitem [{\citenamefont {Liechtenstein}\ \emph {et~al.}(1984)\citenamefont
  {Liechtenstein}, \citenamefont {Katsnelson},\ and\ \citenamefont
  {Gubanov}}]{Liechtenstein1984}%
  \BibitemOpen
  \bibfield  {author} {\bibinfo {author} {\bibfnamefont {A.~I.}\ \bibnamefont
  {Liechtenstein}}, \bibinfo {author} {\bibfnamefont {M.~I.}\ \bibnamefont
  {Katsnelson}},\ and\ \bibinfo {author} {\bibfnamefont {V.~A.}\ \bibnamefont
  {Gubanov}},\ }\bibfield  {title} {\bibinfo {title} {Exchange interactions and
  spin-wave stiffness in ferromagnetic metals},\ }\href
  {https://doi.org/10.1088/0305-4608/14/7/007} {\bibfield  {journal} {\bibinfo
  {journal} {J. Phys. F}\ }\textbf {\bibinfo {volume} {14}},\ \bibinfo {pages}
  {L125} (\bibinfo {year} {1984})}\BibitemShut {NoStop}%
\bibitem [{\citenamefont {Liechtenstein}\ \emph {et~al.}(1987)\citenamefont
  {Liechtenstein}, \citenamefont {Katsnelson}, \citenamefont {Antropov},\ and\
  \citenamefont {Gubanov}}]{Liechtenstein1987}%
  \BibitemOpen
  \bibfield  {author} {\bibinfo {author} {\bibfnamefont {A.~I.}\ \bibnamefont
  {Liechtenstein}}, \bibinfo {author} {\bibfnamefont {M.~I.}\ \bibnamefont
  {Katsnelson}}, \bibinfo {author} {\bibfnamefont {V.~P.}\ \bibnamefont
  {Antropov}},\ and\ \bibinfo {author} {\bibfnamefont {V.~A.}\ \bibnamefont
  {Gubanov}},\ }\bibfield  {title} {\bibinfo {title} {Local spin density
  functional approach to the theory of exchange interactions in ferromagnetic
  metals and alloys},\ }\href
  {https://doi.org/https://doi.org/10.1016/0304-8853(87)90721-9} {\bibfield
  {journal} {\bibinfo  {journal} {J. Magn. Magn. Mater.}\ }\textbf {\bibinfo
  {volume} {67}},\ \bibinfo {pages} {65 } (\bibinfo {year} {1987})}\BibitemShut
  {NoStop}%
\bibitem [{\citenamefont {Turzhevskii}\ \emph {et~al.}(1990)\citenamefont
  {Turzhevskii}, \citenamefont {Likhtenstein},\ and\ \citenamefont
  {Katsnelson}}]{Turzhevskii1990}%
  \BibitemOpen
  \bibfield  {author} {\bibinfo {author} {\bibfnamefont {S.~A.}\ \bibnamefont
  {Turzhevskii}}, \bibinfo {author} {\bibfnamefont {A.~I.}\ \bibnamefont
  {Likhtenstein}},\ and\ \bibinfo {author} {\bibfnamefont {M.~I.}\ \bibnamefont
  {Katsnelson}},\ }\bibfield  {title} {\bibinfo {title} {Degree of localization
  of magnetic moments and the non-heisenberg nature of exchange interactions in
  metals and alloys},\ }\href
  {http://www.mathnet.ru/php/archive.phtml?wshow=paper&jrnid=ftt&paperid=6220&option_lang=eng}
  {\bibfield  {journal} {\bibinfo  {journal} {Sov. Phys. Solid State}\ }\textbf
  {\bibinfo {volume} {32}},\ \bibinfo {pages} {1138} (\bibinfo {year}
  {1990})}\BibitemShut {NoStop}%
\bibitem [{\citenamefont {Secchi}\ \emph {et~al.}(2013)\citenamefont {Secchi},
  \citenamefont {Brener}, \citenamefont {Lichtenstein},\ and\ \citenamefont
  {Katsnelson}}]{Secchi2013}%
  \BibitemOpen
  \bibfield  {author} {\bibinfo {author} {\bibfnamefont {A.}~\bibnamefont
  {Secchi}}, \bibinfo {author} {\bibfnamefont {S.}~\bibnamefont {Brener}},
  \bibinfo {author} {\bibfnamefont {A.}~\bibnamefont {Lichtenstein}},\ and\
  \bibinfo {author} {\bibfnamefont {M.}~\bibnamefont {Katsnelson}},\ }\bibfield
   {title} {\bibinfo {title} {Non-equilibrium magnetic interactions in strongly
  correlated systems},\ }\href
  {https://doi.org/https://doi.org/10.1016/j.aop.2013.03.006} {\bibfield
  {journal} {\bibinfo  {journal} {Ann. Phys. (N. Y.)}\ }\textbf {\bibinfo
  {volume} {333}},\ \bibinfo {pages} {221 } (\bibinfo {year}
  {2013})}\BibitemShut {NoStop}%
\bibitem [{\citenamefont {Secchi}\ \emph {et~al.}(2016)\citenamefont {Secchi},
  \citenamefont {Lichtenstein},\ and\ \citenamefont {Katsnelson}}]{Secchi2016}%
  \BibitemOpen
  \bibfield  {author} {\bibinfo {author} {\bibfnamefont {A.}~\bibnamefont
  {Secchi}}, \bibinfo {author} {\bibfnamefont {A.~I.}\ \bibnamefont
  {Lichtenstein}},\ and\ \bibinfo {author} {\bibfnamefont {M.~I.}\ \bibnamefont
  {Katsnelson}},\ }\bibfield  {title} {\bibinfo {title} {Nonequilibrium
  itinerant-electron magnetism: A time-dependent mean-field theory},\ }\href
  {https://doi.org/10.1103/PhysRevB.94.085153} {\bibfield  {journal} {\bibinfo
  {journal} {Phys. Rev. B}\ }\textbf {\bibinfo {volume} {94}},\ \bibinfo
  {pages} {085153} (\bibinfo {year} {2016})}\BibitemShut {NoStop}%
\bibitem [{\citenamefont {Antropov}\ \emph {et~al.}(1997)\citenamefont
  {Antropov}, \citenamefont {Katsnelson},\ and\ \citenamefont
  {Liechtenstein}}]{Antropov1997}%
  \BibitemOpen
  \bibfield  {author} {\bibinfo {author} {\bibfnamefont {V.}~\bibnamefont
  {Antropov}}, \bibinfo {author} {\bibfnamefont {M.}~\bibnamefont
  {Katsnelson}},\ and\ \bibinfo {author} {\bibfnamefont {A.}~\bibnamefont
  {Liechtenstein}},\ }\bibfield  {title} {\bibinfo {title} {Exchange
  interactions in magnets},\ }\href
  {https://doi.org/10.1016/S0921-4526(97)00203-2} {\bibfield  {journal}
  {\bibinfo  {journal} {Physica B Condens. Matter}\ }\textbf {\bibinfo {volume}
  {237-238}},\ \bibinfo {pages} {336 } (\bibinfo {year} {1997})}\BibitemShut
  {NoStop}%
\bibitem [{\citenamefont {Antropov}\ \emph {et~al.}(1999)\citenamefont
  {Antropov}, \citenamefont {Harmon},\ and\ \citenamefont
  {Smirnov}}]{Antropov1999}%
  \BibitemOpen
  \bibfield  {author} {\bibinfo {author} {\bibfnamefont {V.}~\bibnamefont
  {Antropov}}, \bibinfo {author} {\bibfnamefont {B.}~\bibnamefont {Harmon}},\
  and\ \bibinfo {author} {\bibfnamefont {A.}~\bibnamefont {Smirnov}},\
  }\bibfield  {title} {\bibinfo {title} {Aspects of spin dynamics and magnetic
  interactions},\ }\href {https://doi.org/10.1016/S0304-8853(99)00425-4}
  {\bibfield  {journal} {\bibinfo  {journal} {J. Magn. Magn. Mater.}\ }\textbf
  {\bibinfo {volume} {200}},\ \bibinfo {pages} {148 } (\bibinfo {year}
  {1999})}\BibitemShut {NoStop}%
\bibitem [{\citenamefont {Szilva}\ \emph {et~al.}(2013)\citenamefont {Szilva},
  \citenamefont {Costa}, \citenamefont {Bergman}, \citenamefont {Szunyogh},
  \citenamefont {Nordstr\"om},\ and\ \citenamefont {Eriksson}}]{Szilva2013}%
  \BibitemOpen
  \bibfield  {author} {\bibinfo {author} {\bibfnamefont {A.}~\bibnamefont
  {Szilva}}, \bibinfo {author} {\bibfnamefont {M.}~\bibnamefont {Costa}},
  \bibinfo {author} {\bibfnamefont {A.}~\bibnamefont {Bergman}}, \bibinfo
  {author} {\bibfnamefont {L.}~\bibnamefont {Szunyogh}}, \bibinfo {author}
  {\bibfnamefont {L.}~\bibnamefont {Nordstr\"om}},\ and\ \bibinfo {author}
  {\bibfnamefont {O.}~\bibnamefont {Eriksson}},\ }\bibfield  {title} {\bibinfo
  {title} {{Interatomic Exchange Interactions for Finite-Temperature Magnetism
  and Nonequilibrium Spin Dynamics}},\ }\href
  {https://doi.org/10.1103/PhysRevLett.111.127204} {\bibfield  {journal}
  {\bibinfo  {journal} {Phys. Rev. Lett.}\ }\textbf {\bibinfo {volume} {111}},\
  \bibinfo {pages} {127204} (\bibinfo {year} {2013})}\BibitemShut {NoStop}%
\bibitem [{\citenamefont {Secchi}\ \emph {et~al.}(2015)\citenamefont {Secchi},
  \citenamefont {Lichtenstein},\ and\ \citenamefont {Katsnelson}}]{Secchi2015}%
  \BibitemOpen
  \bibfield  {author} {\bibinfo {author} {\bibfnamefont {A.}~\bibnamefont
  {Secchi}}, \bibinfo {author} {\bibfnamefont {A.}~\bibnamefont
  {Lichtenstein}},\ and\ \bibinfo {author} {\bibfnamefont {M.}~\bibnamefont
  {Katsnelson}},\ }\bibfield  {title} {\bibinfo {title} {Magnetic interactions
  in strongly correlated systems: Spin and orbital contributions},\ }\href
  {https://doi.org/https://doi.org/10.1016/j.aop.2015.05.002} {\bibfield
  {journal} {\bibinfo  {journal} {Ann. Phys. (N. Y.)}\ }\textbf {\bibinfo
  {volume} {360}},\ \bibinfo {pages} {61 } (\bibinfo {year}
  {2015})}\BibitemShut {NoStop}%
\bibitem [{\citenamefont {Szilva}\ \emph {et~al.}(2017)\citenamefont {Szilva},
  \citenamefont {Thonig}, \citenamefont {Bessarab}, \citenamefont {Kvashnin},
  \citenamefont {Rodrigues}, \citenamefont {Cardias}, \citenamefont {Pereiro},
  \citenamefont {Nordstr\"om}, \citenamefont {Bergman}, \citenamefont
  {Klautau},\ and\ \citenamefont {Eriksson}}]{Szilva2017}%
  \BibitemOpen
  \bibfield  {author} {\bibinfo {author} {\bibfnamefont {A.}~\bibnamefont
  {Szilva}}, \bibinfo {author} {\bibfnamefont {D.}~\bibnamefont {Thonig}},
  \bibinfo {author} {\bibfnamefont {P.~F.}\ \bibnamefont {Bessarab}}, \bibinfo
  {author} {\bibfnamefont {Y.~O.}\ \bibnamefont {Kvashnin}}, \bibinfo {author}
  {\bibfnamefont {D.~C.~M.}\ \bibnamefont {Rodrigues}}, \bibinfo {author}
  {\bibfnamefont {R.}~\bibnamefont {Cardias}}, \bibinfo {author} {\bibfnamefont
  {M.}~\bibnamefont {Pereiro}}, \bibinfo {author} {\bibfnamefont
  {L.}~\bibnamefont {Nordstr\"om}}, \bibinfo {author} {\bibfnamefont
  {A.}~\bibnamefont {Bergman}}, \bibinfo {author} {\bibfnamefont {A.~B.}\
  \bibnamefont {Klautau}},\ and\ \bibinfo {author} {\bibfnamefont
  {O.}~\bibnamefont {Eriksson}},\ }\bibfield  {title} {\bibinfo {title}
  {{Theory of noncollinear interactions beyond Heisenberg exchange:
  Applications to bcc Fe}},\ }\href
  {https://doi.org/10.1103/PhysRevB.96.144413} {\bibfield  {journal} {\bibinfo
  {journal} {Phys. Rev. B}\ }\textbf {\bibinfo {volume} {96}},\ \bibinfo
  {pages} {144413} (\bibinfo {year} {2017})}\BibitemShut {NoStop}%
\bibitem [{\citenamefont {Cardias}\ \emph
  {et~al.}(2020{\natexlab{a}})\citenamefont {Cardias}, \citenamefont {Szilva},
  \citenamefont {Bezerra-Neto}, \citenamefont {Ribeiro}, \citenamefont
  {Bergman}, \citenamefont {Kvashnin}, \citenamefont {Fransson}, \citenamefont
  {Klautau}, \citenamefont {Eriksson},\ and\ \citenamefont
  {Nordstr{\"o}m}}]{Cardias2020a}%
  \BibitemOpen
  \bibfield  {author} {\bibinfo {author} {\bibfnamefont {R.}~\bibnamefont
  {Cardias}}, \bibinfo {author} {\bibfnamefont {A.}~\bibnamefont {Szilva}},
  \bibinfo {author} {\bibfnamefont {M.~M.}\ \bibnamefont {Bezerra-Neto}},
  \bibinfo {author} {\bibfnamefont {M.~S.}\ \bibnamefont {Ribeiro}}, \bibinfo
  {author} {\bibfnamefont {A.}~\bibnamefont {Bergman}}, \bibinfo {author}
  {\bibfnamefont {Y.~O.}\ \bibnamefont {Kvashnin}}, \bibinfo {author}
  {\bibfnamefont {J.}~\bibnamefont {Fransson}}, \bibinfo {author}
  {\bibfnamefont {A.~B.}\ \bibnamefont {Klautau}}, \bibinfo {author}
  {\bibfnamefont {O.}~\bibnamefont {Eriksson}},\ and\ \bibinfo {author}
  {\bibfnamefont {L.}~\bibnamefont {Nordstr{\"o}m}},\ }\bibfield  {title}
  {\bibinfo {title} {{First-principles Dzyaloshinskii--Moriya interaction in a
  non-collinear framework}},\ }\href
  {https://doi.org/10.1038/s41598-020-77219-3} {\bibfield  {journal} {\bibinfo
  {journal} {Sci. Rep.}\ }\textbf {\bibinfo {volume} {10}},\ \bibinfo {pages}
  {20339} (\bibinfo {year} {2020}{\natexlab{a}})}\BibitemShut {NoStop}%
\bibitem [{\citenamefont {Cardias}\ \emph
  {et~al.}(2020{\natexlab{b}})\citenamefont {Cardias}, \citenamefont {Bergman},
  \citenamefont {Szilva}, \citenamefont {Kvashnin}, \citenamefont {Fransson},
  \citenamefont {Klautau}, \citenamefont {Eriksson},\ and\ \citenamefont
  {Nordstr\"om}}]{Cardias2020b}%
  \BibitemOpen
  \bibfield  {author} {\bibinfo {author} {\bibfnamefont {R.}~\bibnamefont
  {Cardias}}, \bibinfo {author} {\bibfnamefont {A.}~\bibnamefont {Bergman}},
  \bibinfo {author} {\bibfnamefont {A.}~\bibnamefont {Szilva}}, \bibinfo
  {author} {\bibfnamefont {Y.~O.}\ \bibnamefont {Kvashnin}}, \bibinfo {author}
  {\bibfnamefont {J.}~\bibnamefont {Fransson}}, \bibinfo {author}
  {\bibfnamefont {A.~B.}\ \bibnamefont {Klautau}}, \bibinfo {author}
  {\bibfnamefont {O.}~\bibnamefont {Eriksson}},\ and\ \bibinfo {author}
  {\bibfnamefont {L.}~\bibnamefont {Nordstr\"om}},\ }\href@noop {} {\bibinfo
  {title} {{Dzyaloshinskii-Moriya interaction in absence of spin-orbit
  coupling}}} (\bibinfo {year} {2020}{\natexlab{b}}),\ \Eprint
  {https://arxiv.org/abs/arXiv:2003.04680} {arXiv:2003.04680} \BibitemShut
  {NoStop}%
\bibitem [{\citenamefont {dos Santos~Dias}\ \emph {et~al.}(2021)\citenamefont
  {dos Santos~Dias}, \citenamefont {Brinker}, \citenamefont {L\'aszl\'offy},
  \citenamefont {Ny\'ari}, \citenamefont {Bl\"ugel}, \citenamefont {Szunyogh},\
  and\ \citenamefont {Lounis}}]{Dias2021}%
  \BibitemOpen
  \bibfield  {author} {\bibinfo {author} {\bibfnamefont {M.}~\bibnamefont {dos
  Santos~Dias}}, \bibinfo {author} {\bibfnamefont {S.}~\bibnamefont {Brinker}},
  \bibinfo {author} {\bibfnamefont {A.}~\bibnamefont {L\'aszl\'offy}}, \bibinfo
  {author} {\bibfnamefont {B.}~\bibnamefont {Ny\'ari}}, \bibinfo {author}
  {\bibfnamefont {S.}~\bibnamefont {Bl\"ugel}}, \bibinfo {author}
  {\bibfnamefont {L.}~\bibnamefont {Szunyogh}},\ and\ \bibinfo {author}
  {\bibfnamefont {S.}~\bibnamefont {Lounis}},\ }\bibfield  {title} {\bibinfo
  {title} {Proper and improper chiral magnetic interactions},\ }\href
  {https://doi.org/10.1103/PhysRevB.103.L140408} {\bibfield  {journal}
  {\bibinfo  {journal} {Phys. Rev. B}\ }\textbf {\bibinfo {volume} {103}},\
  \bibinfo {pages} {L140408} (\bibinfo {year} {2021})}\BibitemShut {NoStop}%
\bibitem [{\citenamefont {Bruno}(2003)}]{Bruno2003}%
  \BibitemOpen
  \bibfield  {author} {\bibinfo {author} {\bibfnamefont {P.}~\bibnamefont
  {Bruno}},\ }\bibfield  {title} {\bibinfo {title} {Exchange interaction
  parameters and adiabatic spin-wave spectra of ferromagnets: A ``renormalized
  magnetic force theorem''},\ }\href
  {https://doi.org/10.1103/PhysRevLett.90.087205} {\bibfield  {journal}
  {\bibinfo  {journal} {Phys. Rev. Lett.}\ }\textbf {\bibinfo {volume} {90}},\
  \bibinfo {pages} {087205} (\bibinfo {year} {2003})}\BibitemShut {NoStop}%
\bibitem [{\citenamefont {Antropov}(2003)}]{Antropov2003}%
  \BibitemOpen
  \bibfield  {author} {\bibinfo {author} {\bibfnamefont {V.}~\bibnamefont
  {Antropov}},\ }\bibfield  {title} {\bibinfo {title} {The exchange coupling
  and spin waves in metallic magnets: removal of the long-wave approximation},\
  }\href {https://doi.org/https://doi.org/10.1016/S0304-8853(03)00206-3}
  {\bibfield  {journal} {\bibinfo  {journal} {J. Magn. Magn. Mater.}\ }\textbf
  {\bibinfo {volume} {262}},\ \bibinfo {pages} {L192 } (\bibinfo {year}
  {2003})}\BibitemShut {NoStop}%
\bibitem [{\citenamefont {Katsnelson}\ and\ \citenamefont
  {Lichtenstein}(2004)}]{Katsnelson2004}%
  \BibitemOpen
  \bibfield  {author} {\bibinfo {author} {\bibfnamefont {M.~I.}\ \bibnamefont
  {Katsnelson}}\ and\ \bibinfo {author} {\bibfnamefont {A.~I.}\ \bibnamefont
  {Lichtenstein}},\ }\bibfield  {title} {\bibinfo {title} {Magnetic
  susceptibility, exchange interactions and spin-wave spectra in the local spin
  density approximation},\ }\href {https://doi.org/10.1088/0953-8984/16/41/023}
  {\bibfield  {journal} {\bibinfo  {journal} {J. Phys. Condens. Matter}\
  }\textbf {\bibinfo {volume} {16}},\ \bibinfo {pages} {7439} (\bibinfo {year}
  {2004})}\BibitemShut {NoStop}%
\bibitem [{\citenamefont {Stocks}\ \emph {et~al.}(1998)\citenamefont {Stocks},
  \citenamefont {Ujfalussy}, \citenamefont {Wang}, \citenamefont {Nicholson},
  \citenamefont {Shelton}, \citenamefont {Wang}, \citenamefont {Canning},\ and\
  \citenamefont {Gy\"orffy}}]{Stocks1998}%
  \BibitemOpen
  \bibfield  {author} {\bibinfo {author} {\bibfnamefont {G.~M.}\ \bibnamefont
  {Stocks}}, \bibinfo {author} {\bibfnamefont {B.}~\bibnamefont {Ujfalussy}},
  \bibinfo {author} {\bibfnamefont {X.}~\bibnamefont {Wang}}, \bibinfo {author}
  {\bibfnamefont {D.~M.~C.}\ \bibnamefont {Nicholson}}, \bibinfo {author}
  {\bibfnamefont {W.~A.}\ \bibnamefont {Shelton}}, \bibinfo {author}
  {\bibfnamefont {Y.}~\bibnamefont {Wang}}, \bibinfo {author} {\bibfnamefont
  {A.}~\bibnamefont {Canning}},\ and\ \bibinfo {author} {\bibfnamefont {B.~L.}\
  \bibnamefont {Gy\"orffy}},\ }\bibfield  {title} {\bibinfo {title} {Towards a
  constrained local moment model for first principles spin dynamics},\ }\href
  {https://doi.org/10.1080/13642819808206775} {\bibfield  {journal} {\bibinfo
  {journal} {Philos. Mag. B}\ }\textbf {\bibinfo {volume} {78}},\ \bibinfo
  {pages} {665} (\bibinfo {year} {1998})}\BibitemShut {NoStop}%
\bibitem [{\citenamefont {Ujfalussy}\ \emph {et~al.}(1999)\citenamefont
  {Ujfalussy}, \citenamefont {Wang}, \citenamefont {Nicholson}, \citenamefont
  {Shelton}, \citenamefont {Stocks}, \citenamefont {Wang},\ and\ \citenamefont
  {Gyorffy}}]{Ujfalussy1999}%
  \BibitemOpen
  \bibfield  {author} {\bibinfo {author} {\bibfnamefont {B.}~\bibnamefont
  {Ujfalussy}}, \bibinfo {author} {\bibfnamefont {X.-D.}\ \bibnamefont {Wang}},
  \bibinfo {author} {\bibfnamefont {D.~M.~C.}\ \bibnamefont {Nicholson}},
  \bibinfo {author} {\bibfnamefont {W.~A.}\ \bibnamefont {Shelton}}, \bibinfo
  {author} {\bibfnamefont {G.~M.}\ \bibnamefont {Stocks}}, \bibinfo {author}
  {\bibfnamefont {Y.}~\bibnamefont {Wang}},\ and\ \bibinfo {author}
  {\bibfnamefont {B.~L.}\ \bibnamefont {Gyorffy}},\ }\bibfield  {title}
  {\bibinfo {title} {Constrained density functional theory for first principles
  spin dynamics},\ }\href {https://doi.org/10.1063/1.370494} {\bibfield
  {journal} {\bibinfo  {journal} {J. Appl. Phys.}\ }\textbf {\bibinfo {volume}
  {85}},\ \bibinfo {pages} {4824} (\bibinfo {year} {1999})}\BibitemShut
  {NoStop}%
\bibitem [{\citenamefont {Streib}\ \emph {et~al.}(2020)\citenamefont {Streib},
  \citenamefont {Borisov}, \citenamefont {Pereiro}, \citenamefont {Bergman},
  \citenamefont {Sj\"oqvist}, \citenamefont {Delin}, \citenamefont {Eriksson},\
  and\ \citenamefont {Thonig}}]{Streib2020}%
  \BibitemOpen
  \bibfield  {author} {\bibinfo {author} {\bibfnamefont {S.}~\bibnamefont
  {Streib}}, \bibinfo {author} {\bibfnamefont {V.}~\bibnamefont {Borisov}},
  \bibinfo {author} {\bibfnamefont {M.}~\bibnamefont {Pereiro}}, \bibinfo
  {author} {\bibfnamefont {A.}~\bibnamefont {Bergman}}, \bibinfo {author}
  {\bibfnamefont {E.}~\bibnamefont {Sj\"oqvist}}, \bibinfo {author}
  {\bibfnamefont {A.}~\bibnamefont {Delin}}, \bibinfo {author} {\bibfnamefont
  {O.}~\bibnamefont {Eriksson}},\ and\ \bibinfo {author} {\bibfnamefont
  {D.}~\bibnamefont {Thonig}},\ }\bibfield  {title} {\bibinfo {title} {Equation
  of motion and the constraining field in \textit{ab initio} spin dynamics},\
  }\href {https://doi.org/10.1103/PhysRevB.102.214407} {\bibfield  {journal}
  {\bibinfo  {journal} {Phys. Rev. B}\ }\textbf {\bibinfo {volume} {102}},\
  \bibinfo {pages} {214407} (\bibinfo {year} {2020})}\BibitemShut {NoStop}%
\bibitem [{\citenamefont {de~Almeida}\ \emph {et~al.}(2021)\citenamefont
  {de~Almeida}, \citenamefont {Barreteau}, \citenamefont {Thibaudeau},\ and\
  \citenamefont {Fu}}]{Cardias2021}%
  \BibitemOpen
  \bibfield  {author} {\bibinfo {author} {\bibfnamefont {R.~C.~A.}\
  \bibnamefont {de~Almeida}}, \bibinfo {author} {\bibfnamefont
  {C.}~\bibnamefont {Barreteau}}, \bibinfo {author} {\bibfnamefont
  {P.}~\bibnamefont {Thibaudeau}},\ and\ \bibinfo {author} {\bibfnamefont
  {C.~C.}\ \bibnamefont {Fu}},\ }\href@noop {} {\bibinfo {title} {Spin dynamics
  from a constrained magnetic tight-binding model}} (\bibinfo {year} {2021}),\
  \Eprint {https://arxiv.org/abs/arXiv:2101.06121} {arXiv:2101.06121}
  \BibitemShut {NoStop}%
\bibitem [{\citenamefont {Solovyev}(2021)}]{Solovyev2020}%
  \BibitemOpen
  \bibfield  {author} {\bibinfo {author} {\bibfnamefont {I.~V.}\ \bibnamefont
  {Solovyev}},\ }\bibfield  {title} {\bibinfo {title} {Exchange interactions
  and magnetic force theorem},\ }\href
  {https://doi.org/10.1103/PhysRevB.103.104428} {\bibfield  {journal} {\bibinfo
   {journal} {Phys. Rev. B}\ }\textbf {\bibinfo {volume} {103}},\ \bibinfo
  {pages} {104428} (\bibinfo {year} {2021})}\BibitemShut {NoStop}%
\bibitem [{\citenamefont {Jacobsson}\ \emph {et~al.}(2017)\citenamefont
  {Jacobsson}, \citenamefont {Johansson}, \citenamefont {Gorbatov},
  \citenamefont {Le\v{z}ai\'{c}}, \citenamefont {Sanyal}, \citenamefont
  {Bl\"ugel},\ and\ \citenamefont {Etz}}]{Jacobsson2017}%
  \BibitemOpen
  \bibfield  {author} {\bibinfo {author} {\bibfnamefont {A.}~\bibnamefont
  {Jacobsson}}, \bibinfo {author} {\bibfnamefont {G.}~\bibnamefont
  {Johansson}}, \bibinfo {author} {\bibfnamefont {O.~I.}\ \bibnamefont
  {Gorbatov}}, \bibinfo {author} {\bibfnamefont {M.}~\bibnamefont
  {Le\v{z}ai\'{c}}}, \bibinfo {author} {\bibfnamefont {B.}~\bibnamefont
  {Sanyal}}, \bibinfo {author} {\bibfnamefont {S.}~\bibnamefont {Bl\"ugel}},\
  and\ \bibinfo {author} {\bibfnamefont {C.}~\bibnamefont {Etz}},\ }\href@noop
  {} {\bibinfo {title} {Parameterisation of non-collinear energy landscapes in
  itinerant magnets}} (\bibinfo {year} {2017}),\ \Eprint
  {https://arxiv.org/abs/arXiv:1702.00599} {arXiv:1702.00599} \BibitemShut
  {NoStop}%
\bibitem [{\citenamefont {Halilov}\ \emph {et~al.}(1998)\citenamefont
  {Halilov}, \citenamefont {Eschrig}, \citenamefont {Perlov},\ and\
  \citenamefont {Oppeneer}}]{Halilov1998}%
  \BibitemOpen
  \bibfield  {author} {\bibinfo {author} {\bibfnamefont {S.~V.}\ \bibnamefont
  {Halilov}}, \bibinfo {author} {\bibfnamefont {H.}~\bibnamefont {Eschrig}},
  \bibinfo {author} {\bibfnamefont {A.~Y.}\ \bibnamefont {Perlov}},\ and\
  \bibinfo {author} {\bibfnamefont {P.~M.}\ \bibnamefont {Oppeneer}},\
  }\bibfield  {title} {\bibinfo {title} {{Adiabatic spin dynamics from
  spin-density-functional theory: Application to Fe, Co, and Ni}},\ }\href
  {https://doi.org/10.1103/PhysRevB.58.293} {\bibfield  {journal} {\bibinfo
  {journal} {Phys. Rev. B}\ }\textbf {\bibinfo {volume} {58}},\ \bibinfo
  {pages} {293} (\bibinfo {year} {1998})}\BibitemShut {NoStop}%
\bibitem [{\citenamefont {Kohn}\ and\ \citenamefont {Sham}(1965)}]{Kohn1965}%
  \BibitemOpen
  \bibfield  {author} {\bibinfo {author} {\bibfnamefont {W.}~\bibnamefont
  {Kohn}}\ and\ \bibinfo {author} {\bibfnamefont {L.~J.}\ \bibnamefont
  {Sham}},\ }\bibfield  {title} {\bibinfo {title} {Self-consistent equations
  including exchange and correlation effects},\ }\href
  {https://doi.org/10.1103/PhysRev.140.A1133} {\bibfield  {journal} {\bibinfo
  {journal} {Phys. Rev.}\ }\textbf {\bibinfo {volume} {140}},\ \bibinfo {pages}
  {A1133} (\bibinfo {year} {1965})}\BibitemShut {NoStop}%
\bibitem [{\citenamefont {Brooks}\ and\ \citenamefont
  {Johansson}(1983)}]{Brooks1983}%
  \BibitemOpen
  \bibfield  {author} {\bibinfo {author} {\bibfnamefont {M.~S.~S.}\
  \bibnamefont {Brooks}}\ and\ \bibinfo {author} {\bibfnamefont
  {B.}~\bibnamefont {Johansson}},\ }\bibfield  {title} {\bibinfo {title}
  {Exchange integral matrices and cohesive energies of transition metal
  atoms},\ }\href {https://doi.org/10.1088/0305-4608/13/10/003} {\bibfield
  {journal} {\bibinfo  {journal} {J. Phys. F: Met. Phys.}\ }\textbf {\bibinfo
  {volume} {13}},\ \bibinfo {pages} {L197} (\bibinfo {year}
  {1983})}\BibitemShut {NoStop}%
\bibitem [{\citenamefont {Aut{\`{e}}s}\ \emph {et~al.}(2006)\citenamefont
  {Aut{\`{e}}s}, \citenamefont {Barreteau}, \citenamefont {Spanjaard},\ and\
  \citenamefont {Desjonqu{\`{e}}res}}]{Aute2006}%
  \BibitemOpen
  \bibfield  {author} {\bibinfo {author} {\bibfnamefont {G.}~\bibnamefont
  {Aut{\`{e}}s}}, \bibinfo {author} {\bibfnamefont {C.}~\bibnamefont
  {Barreteau}}, \bibinfo {author} {\bibfnamefont {D.}~\bibnamefont
  {Spanjaard}},\ and\ \bibinfo {author} {\bibfnamefont {M.-C.}\ \bibnamefont
  {Desjonqu{\`{e}}res}},\ }\bibfield  {title} {\bibinfo {title} {Magnetism of
  iron: from the bulk to the monatomic wire},\ }\href
  {https://doi.org/10.1088/0953-8984/18/29/018} {\bibfield  {journal} {\bibinfo
   {journal} {J. Phys. Condens. Matter}\ }\textbf {\bibinfo {volume} {18}},\
  \bibinfo {pages} {6785} (\bibinfo {year} {2006})}\BibitemShut {NoStop}%
\bibitem [{\citenamefont {Schena}(2010)}]{Schena2010}%
  \BibitemOpen
  \bibfield  {author} {\bibinfo {author} {\bibfnamefont {T.}~\bibnamefont
  {Schena}},\ }\emph {\bibinfo {title} {{Tight-Binding Treatment of Complex
  Magnetic Structures in Low-Dimensional Systems}}},\ \href
  {https://www.fz-juelich.de/SharedDocs/Downloads/PGI/PGI-1/EN/Schena_diploma_pdf.pdf}
  {\bibinfo {type} {Diploma thesis}},\ \bibinfo  {school} {TH Aachen} (\bibinfo
  {year} {2010})\BibitemShut {NoStop}%
\bibitem [{\citenamefont {Katsnelson}\ and\ \citenamefont
  {Lichtenstein}(2000)}]{Katsnelson2000}%
  \BibitemOpen
  \bibfield  {author} {\bibinfo {author} {\bibfnamefont {M.~I.}\ \bibnamefont
  {Katsnelson}}\ and\ \bibinfo {author} {\bibfnamefont {A.~I.}\ \bibnamefont
  {Lichtenstein}},\ }\bibfield  {title} {\bibinfo {title} {First-principles
  calculations of magnetic interactions in correlated systems},\ }\href
  {https://doi.org/10.1103/PhysRevB.61.8906} {\bibfield  {journal} {\bibinfo
  {journal} {Phys. Rev. B}\ }\textbf {\bibinfo {volume} {61}},\ \bibinfo
  {pages} {8906} (\bibinfo {year} {2000})}\BibitemShut {NoStop}%
\bibitem [{\citenamefont {Nomoto}\ \emph {et~al.}(2020)\citenamefont {Nomoto},
  \citenamefont {Koretsune},\ and\ \citenamefont {Arita}}]{Nomoto2020}%
  \BibitemOpen
  \bibfield  {author} {\bibinfo {author} {\bibfnamefont {T.}~\bibnamefont
  {Nomoto}}, \bibinfo {author} {\bibfnamefont {T.}~\bibnamefont {Koretsune}},\
  and\ \bibinfo {author} {\bibfnamefont {R.}~\bibnamefont {Arita}},\ }\bibfield
   {title} {\bibinfo {title} {Local force method for the ab initio
  tight-binding model: Effect of spin-dependent hopping on exchange
  interactions},\ }\href {https://doi.org/10.1103/PhysRevB.102.014444}
  {\bibfield  {journal} {\bibinfo  {journal} {Phys. Rev. B}\ }\textbf {\bibinfo
  {volume} {102}},\ \bibinfo {pages} {014444} (\bibinfo {year}
  {2020})}\BibitemShut {NoStop}%
\bibitem [{CAH()}]{CAHMD}%
  \BibitemOpen
  \href@noop {} {}\bibinfo {note} {{Computer code CAHMD, classical atomistic
  Heisenberg magnetization dynamics. A computer program package for atomistic
  magnetization dynamics simulations. (Danny Thonig, danny.thonig@oru.se, 2013)
  (unpublished, available from
  \url{https://cahmd.gitlab.io/cahmdweb/})}}\BibitemShut {NoStop}%
\bibitem [{\citenamefont {Slater}\ and\ \citenamefont
  {Koster}(1954)}]{Slater1954}%
  \BibitemOpen
  \bibfield  {author} {\bibinfo {author} {\bibfnamefont {J.~C.}\ \bibnamefont
  {Slater}}\ and\ \bibinfo {author} {\bibfnamefont {G.~F.}\ \bibnamefont
  {Koster}},\ }\bibfield  {title} {\bibinfo {title} {{Simplified LCAO Method
  for the Periodic Potential Problem}},\ }\href
  {https://doi.org/10.1103/PhysRev.94.1498} {\bibfield  {journal} {\bibinfo
  {journal} {Phys. Rev.}\ }\textbf {\bibinfo {volume} {94}},\ \bibinfo {pages}
  {1498} (\bibinfo {year} {1954})}\BibitemShut {NoStop}%
\bibitem [{\citenamefont {Thonig}\ and\ \citenamefont
  {Henk}(2014)}]{Thonig2014}%
  \BibitemOpen
  \bibfield  {author} {\bibinfo {author} {\bibfnamefont {D.}~\bibnamefont
  {Thonig}}\ and\ \bibinfo {author} {\bibfnamefont {J.}~\bibnamefont {Henk}},\
  }\bibfield  {title} {\bibinfo {title} {{Gilbert damping tensor within the
  breathing Fermi surface model: anisotropy and non-locality}},\ }\href
  {https://doi.org/10.1088/1367-2630/16/1/013032} {\bibfield  {journal}
  {\bibinfo  {journal} {New J. Phys}\ }\textbf {\bibinfo {volume} {16}},\
  \bibinfo {pages} {013032} (\bibinfo {year} {2014})}\BibitemShut {NoStop}%
\bibitem [{\citenamefont {Mankovsky}\ \emph {et~al.}(2020)\citenamefont
  {Mankovsky}, \citenamefont {Polesya},\ and\ \citenamefont
  {Ebert}}]{Mankovsky2020b}%
  \BibitemOpen
  \bibfield  {author} {\bibinfo {author} {\bibfnamefont {S.}~\bibnamefont
  {Mankovsky}}, \bibinfo {author} {\bibfnamefont {S.}~\bibnamefont {Polesya}},\
  and\ \bibinfo {author} {\bibfnamefont {H.}~\bibnamefont {Ebert}},\ }\bibfield
   {title} {\bibinfo {title} {Exchange coupling constants at finite
  temperature},\ }\href {https://doi.org/10.1103/PhysRevB.102.134434}
  {\bibfield  {journal} {\bibinfo  {journal} {Phys. Rev. B}\ }\textbf {\bibinfo
  {volume} {102}},\ \bibinfo {pages} {134434} (\bibinfo {year}
  {2020})}\BibitemShut {NoStop}%
\bibitem [{\citenamefont {Malki}\ and\ \citenamefont
  {Uhrig}(2020)}]{Malki2020}%
  \BibitemOpen
  \bibfield  {author} {\bibinfo {author} {\bibfnamefont {M.}~\bibnamefont
  {Malki}}\ and\ \bibinfo {author} {\bibfnamefont {G.~S.}\ \bibnamefont
  {Uhrig}},\ }\bibfield  {title} {\bibinfo {title} {Topological magnetic
  excitations},\ }\href {https://doi.org/10.1209/0295-5075/132/20003}
  {\bibfield  {journal} {\bibinfo  {journal} {{EPL}}\ }\textbf {\bibinfo
  {volume} {132}},\ \bibinfo {pages} {20003} (\bibinfo {year}
  {2020})}\BibitemShut {NoStop}%
\end{thebibliography}
\end{document}